\documentclass[aps,prd,showpacs]{revtex4-1}
\usepackage{amsmath} 
\usepackage{amssymb} 
\usepackage{graphics}
\usepackage{epsfig}
\usepackage{slashed}

\def\bge{\begin{equation}}
\def\ene{\end{equation}}
\def\bgea{\begin{eqnarray}}
\def\enea{\end{eqnarray}}

\usepackage{longtable}
\usepackage{latexsym}

\usepackage{amssymb}
\usepackage{amsfonts} 
\usepackage{amsmath}

\begin{document}

\title{THE TRIPLET PHOTOPRODUCTION ON A FREE ELECTRON AS A POSSIBLE WAY TO SEARCH FOR A DARK PHOTON. }

\vspace{4mm}

\author{G.I.~Gakh}

\affiliation{\it NSC ''Kharkov Institute of Physics and Technology'',
Akademicheskaya, 1, 61108 Kharkov,  and \\
V.N.~Karazin Kharkiv National University, 61022 Kharkov, Ukraine}

\author{M.I.~Konchatnij}
\affiliation{\it NSC ''Kharkov Institute of Physics and Technology'',
Akademicheskaya, 1, 61108 Kharkov, and \\
V.N.~Karazin Kharkiv National University, 61022 Kharkov, Ukraine}

\author{N.P.~ Merenkov}
\affiliation{\it NSC ''Kharkov Institute of Physics and Technology'',
Akademicheskaya, 1, 61108 Kharkov,  and \\
 V.N.~Karazin Kharkiv National University, 61022 Kharkov, Ukraine}

\date{\today}
\vspace{0.3cm}

\begin{abstract}
The process of the triplet production on a free electron,
$\gamma e^-\to e^+e^-e^-$, has been investigated as a reaction
where a dark photon, $A'$, is produced as a virtual state with subsequent decay
into a $e^+e^-$-pair.
This effect arises due to the so-called kinetic mixing and is characterized by the small parameter
$\epsilon$ describing the coupling
strength relative to the electric charge e.

The search of $A'$ in this process has advantage because the
background to the $A'$ signal is pure QED. This QED background is described
by eight Feynman diagrams taking
into account the identity of final electrons. As concern $A'$, we leave its contribution
in Compton-like diagrams only since, in this case,
the virtual dark photon is time-like and its propagator has the Breit-Wigner form. So, near the resonance $A'$ can manifest itself.
 The contribution of $A'$ in
 Borsellino diagrams is negligible since, in this case, the
virtual dark photon is space-like, the $A'$ propagator does not
peak and effect is proportional at least to $\epsilon^2$.
We calculate the
distributions over the invariant masses of the both produced $e^+e^-$ pairs
and search for the kinematical region
where the Compton-like diagrams contribution is not suppressed as compared with the Borsellino one.
 We estimate what value of the parameter $\epsilon$,
as a function of the dark photon mass, can be obtained at given
number of the measured events.
\end{abstract}

\maketitle

\section{Introduction}

As is known, some experimental discoveries, such as the neutrino
oscillation, the existence of the dark matter (its nature and
interaction are unknown today) and some astrophysical data, lead to
the necessity of the consideration the physics beyond the SM (see
the reviews \cite{Rev12, Rev13, Rev15, Rev16}). One of the possible
new particle is the so-called dark photon, $A'$. It is massive vector
boson that can mix with the ordinary photon via "kinetic-mixing"
\cite{H86}. Its mass and interaction strength are not predicted
unambiguously by the theory since the mass of the dark photon can
arise via different mechanisms. Various theoretically motivated
regions of the dark photon mass are shown in Figs. 6-2, 6-3 in the
Ref. \cite{Rev12}. For the dark photon with mass larger than 1 MeV,
it is possible its production in electron (proton) fixed-target
experiments or at hadron or electron-positron colliders (see the
references in review \cite{Rev12}).

The experimental investigation of the dark photon effects are
planned or performed in various laboratories: APEX \cite{ESTW11,
ABet11}, HPS \cite{HPS}, DarkLight \cite{FOT10} (JLAB), MAMI
\cite{Met11} (fixed-target experiments) and VEPP-3 \cite{W09}
(electron-positron collider). The manifestation of the dark photon
was searched also in the decay of the known particles. The authors of
Ref. \cite{Fet03} have studied radiative pion decays $\pi^ +
\rightarrow e^+\nu\gamma$. The measurements were performed in the
$\pi$E1 channel at the Paul Scherrer Institute (PSI), Switzerland.
The dark photon was searched in the decay of $\pi^0$-meson
($\pi^0\rightarrow \gamma A'\rightarrow \gamma e^+e^-$)\cite{Xet13}
which produced in proton nuclei collisions at HIAF facility (China).
The decay of $\pi^0$-meson was also used to search for the dark photon
in WASA-at-COSY experiment (J\"{u}lich, Germany) \cite{M14} ($\pi^0$-mesons were produced in the reaction
$pp\rightarrow pp\pi^0$) and at CERN ($\pi^0$-mesons were produced
in the decays of K-mesons, $K^{\pm}\rightarrow \pi^{\pm}\pi^0$) \cite{G14}.
A search for a dark photon signal in inclusive dielectron spectra in
proton-induced reactions on either a liquid hydrogen target or
nuclei performed at the GSI in Darmstadt \cite{Aet13}. An upper
limit on the dark photon mixing parameter in the mass range m(A')=
0.02 - 0.6 GeV/c$^2$ has been established. The current status of the
limits on the dark photons parameters from electron beam dump experiments is
summarized in \cite{A12}. It has been demonstrated at JLab that
electron-beam fixed-target experiments would have powerful discovery
potential for a dark matter in the MeV-GeV mass range \cite{IKST14}.

Theoretically, the production of the dark photon in various
reactions has been investigated in a number of papers.
Bjorken et al. \cite{BEST09} discussed several possible
experimental schemes for the search for a $A'$ in the most likely mass
range of a few MeV/c$^2$ up to a few GeV/c$^2$. They stated that the
fixed-target experiments are ideally suited for discovering
few MeV-GeV mass dark photon.
The
production of the dark photon in the process of the electron
scattering on the proton or heavy nucleus has been investigated in
Ref. \cite{BMV13} (\cite{BV14}) for the experimental conditions of
the MAMI (JLab) experiment \cite{Met11} (\cite{ABet11}). The authors
of Ref. \cite{CT16} proposed to use rare leptonic decays of kaons
and pions, $K^+(\pi^+)\rightarrow \mu^+\nu_{\mu}e^+e^-$, to study the
light dark photon (with mass about 10 MeV). The constraints on the
dark photon in the 0.01 - 100 keV mass range are derived in Ref.
\cite{APPR14} (indirect constraints from $A'\rightarrow 3\gamma $
decay are also revisited). The proposal to search for light dark
photon using the Compton-like process, $\gamma e\rightarrow A'e$, in
a nuclear reactor was suggested in Ref. \cite{P17}. Using the
existed experimental data, the limits on the kinetic mixing
parameter were derived. Some results on the phenomenology of the
dark photon in the mass range of a few MeV to GeV have been
presented in \cite{P09}, where g-2 of muons and electrons together
with other precision QED data, as well as radiative decays of
strange particles, were analyzed.

At proposed project IRIDE (Frascati, Italy) \cite{IRIDE}, the
physical program consider a search for dark photon via the lepton
triplet production process in the electron-photon collision. The
main QED process of the lepton triplet production is determined by
Bethe-Heitler diagrams and the virtual Compton scattering diagrams.
Therefore, the virtual Compton scattering part of the QED process is
of the most importance, as it is intimately related to the dark
photon production, while the Bethe-Heitler contribution must be
reduced as much as possible. This can be done by specific angular
criteria for the event selection.

In this paper, we consider the lepton triplet production in the
photon scattering by electrons. The dark photon production is taken
into account only in the Compton-like diagrams where
electron-positron pair from the decay $A'\rightarrow e^+e^-$ can be
in resonance. The advantage of this consideration is that the
background, in this case, is a pure QED process $\gamma
e^-\rightarrow e^+e^-e^-$ which can be calculated with the
required precision. We calculate the distributions over the
invariant masses of the produced $e^+e^-$ pairs taking into account
the identity of the final electrons. All eight Feynman diagrams,
that determine the background reaction, was taken into account when
calculating distributions in $\gamma e^-\rightarrow e^+e^-e^-$.
We estimate what value of the parameter $\epsilon$,
as a function of the dark photon mass, can be obtained at given
number of the measured events.

In Sec. II the formalism of the calculation of the distribution
is given. Sec. II\,A is devoted to the description of the kinematical
variables. The calculation of the double differential distribution
over the masses of the produced $e^+e^-$ pairs, caused by the QED
mechanism, is given in Sec. II\,B. The dark photon contribution to the
distribution is calculated in Sec. II\,C. Sec. III is devoted to the
analysis to the dark photon effects in the considered reaction and
to the estimation of the parameter $\epsilon $ as a function of the
dark photon mass and the number of the measured events.
The detailed analysis of the kinematics for the restricted phase space is given in Appendix A.
The Appendix B contains the analytical
expressions for the contributions into distribution over invariant masses of $e^+e^--$pairs which are caused by the separate diagrams and their
interferences in the case of nonrestricted phase space.


\section{Formalism}

As we mentioned in the Introduction, the dark photon can manifest
itself as some resonance state decaying in the electron-positron
pair. In
this case, the process of the triplet production,
 \bge\label{eq:1}
\gamma(k)+e^-(p)\rightarrow e^+(p_3)+e^-(p_1)+e^-(p_2)\,, \ene 
can
be used to search for the dark photon signal due to the creation of
two electron-positron systems, with the invariant masses squared
$$s_1=(p_3+p_1)^2 \,, \ \ s_2=(p_3+p_2)^2,$$
through the dark photon intermediate state. The double differential
distribution over the $s_1$ and $s_2$ variables, in the process
(\ref{eq:1}), is the most suitable for this goal since it takes into
account the identity of the final electrons.

The pure QED distribution is the background which exceeds
significantly the dark photon effect. Thus, it has to be calculated
as precise as possible and accounted for in the searches for the
dark photon signal. The QED amplitude, for the triplet production
process, is described by eight diagrams, four of them are presented
in Fig.\,1, and the another four can be derived from these diagrams
by permutation the electron 4-momenta $(p_1\leftrightarrows p_2).$

\begin{figure}
\includegraphics[width=0.44\textwidth]{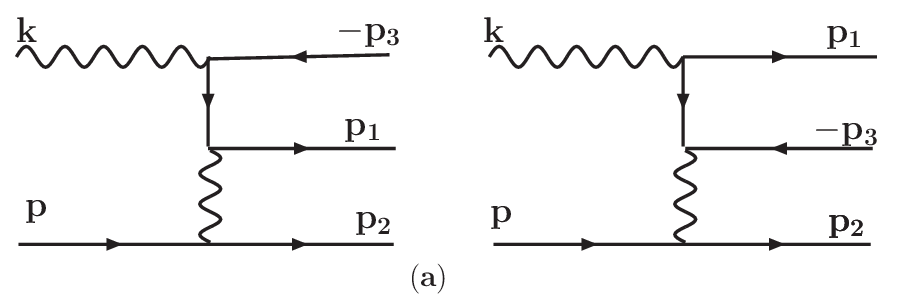}
\hspace{0.4cm}
\includegraphics[width=0.44\textwidth]{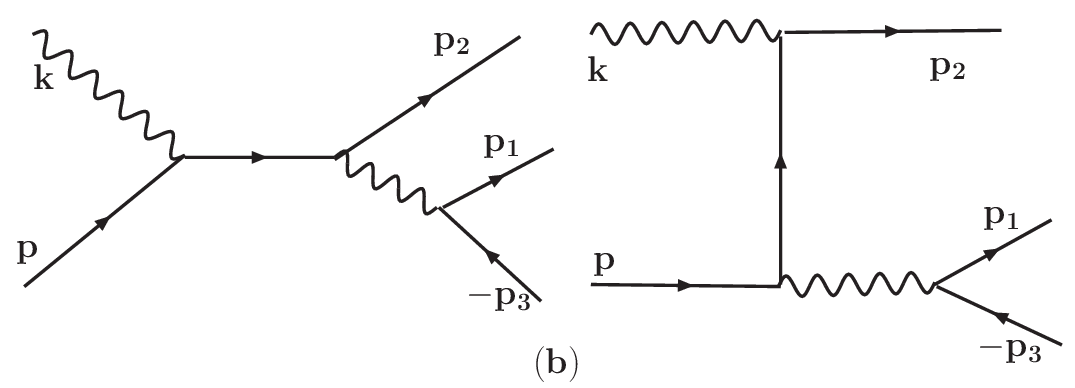}
\caption{The Feynman diagrams for the QED amplitude of the process
(\ref{eq:1}). The diagrams ({\it a}) are the so-called Borsellino
diagrams and the diagrams ({\it b}) are the Compton-like ones. To
take into account the final electron identity, one has to add
corresponding diagrams with permutation of the 4-momenta $p_1$ and
$p_2$.}
\end{figure}

In present paper, we calculate the double differential distribution
over the $s_1$ and $s_2$  variables to calculate the QED background
and to find the regions of the $s_1$ and $s_2$ where the single
photon amplitudes (Compton-like) contribute  on the level of the
two photon amplitudes (Borsellino) and even more.

\subsection{Kinematics}

The process (1) is the $2\to 3$ one, and for its description we can
use a few sets of the kinematical variables \cite{BK1973}. To obtain
the distribution over the $s_1$ and $s_2$ variables, we use the
description in terms of five invariants as follows
$$s=(k+p)^2=(p_1+p_2+p_3)^2\,, \ s_1=(p_1+p_3)^2=(k+p-p_2)^2\,,
\ \ s_2=(p_2+p_3)^2=(k+p-p_1)^2\,,$$
\bge\label{eq:2}
t_1=(k-p_1)^2=(p_2+p_3-p)^2\,,\ \ t_2=(p-p_2)^2=(p+p_3-k)^2\,.
\ene
The scalar products of the 4-momenta in the reaction are expressed in
terms of these invariants
$$2(k\,p)=s-m^2\,, \ 2(p_1\,p_3)=s_1-2m^2\,, \ 2(p_2\,p_3)=s_2-2m^2\,,
\ 2(k\,p_1)=m^2-t_1\,, $$
\bge\label{eq:3}
2(p\,p_2)=2m^2-t_2\,, \ 2(k\,p_2)=s-s_1+t_2-m^2\,, \ 2(k\,p_3)=s_1+t_1-t_2-m^2\,,
\ene
$$2(p_1\,p_2)=s-s_1-s_2+m^2\,, \ 2(p\,p_1)=s-s_2+t_1\,, \ 2(p\,p_3)=s_2-t_1+t_2-m^2\,.$$

Bearing in mind the azimuthal symmetry relative to the photon beam
direction, the phase space of the final particles in (1) can be written
as \cite{BK1973}
\bge\label{eq:4}
d\,R_3=\frac{d^3p_1}{2\,E_1}\,\frac{d^3p_2}{2\,E_2}\,\frac{d^3p_3}{2\,E_3}\,\delta(k+p-p_1-p_2-p_3)=\frac{\pi}{16(s-m^2)}
\frac{dt_1\,dt_2\,ds_1\,ds_2}{\sqrt{-\Delta}}\,,
\ene
where $\Delta$ is the Gramian determinant. In terms of the used variables,
it can be written as
$$\Delta=\frac{1}{16}\left|\begin{array}{c c c c} ~0~~~~~~~s-m^2~~~~~~~~~m^2-t_1~~~~~~~~~~s-s_1+t_2-m^2~~~~\\
s-m^2~~~~~~~2m^2~~~~~~~~~~s-s_2+t_1~~~~~~~~~2m^2-t_2~~~~\\
m^2-t_1~~~~~s-s_2+t_1~~~~~~2m^2~~~~~~~s-s_1-s_2+m^2~~~~\\
s-s_1+t_2-m^2~~~2m^2-t_2~~~~s-s_1-s_2+m^2~~~2m^2\end{array}\right|\,.$$

To derive the studied distribution, we have to integrate the
differential cross section with respect to the variables $t_1$ and
$t_2.$ The limits of integration can be obtained from the condition
of the positivity of $(-\Delta).$ Solving equation $\Delta=0$ with
respect to the variable $t_1,$ we obtain \bge\label{eq:5}
t_{1-}<t_1<t_{1+}\,, \ \ \ t_{1\pm}=\frac{A\pm
2\sqrt{B}}{(s-s_1)^2-2(s+s_1)m^2+m^4}\,, \ene
$$A=s_1s_2(s_1-t_2)+s^2t_2-s[s_2t_2+s_1(s_2+t_2)]+
m^2[s^2+(s_1+s_2)(s_1+t_2)]-m^4(2s+4s_1+t_2)+m^6\,,$$
$$B=\big[st_2(s-s_1+t_2)+m^2(s_1^2-2st_2-s_1t_2)+
m^4t_2\big]\big[s_1s_2(s_1+s_2-s)+m^2(s^2-3s_1s_2)-2m^2s+m^6\big]\,.$$

The limits of the second integration over the variable $t_2$, at fixed
$s_1$ and $s_2$, are determined as the roots of the first factor in
the expression B, namely
\bge\label{eq:6}
t_{2-}\,<t_2<t_{2+}\,, \ \ \ t_{2\pm}=\frac{C\pm (s-m^2)\lambda_1}{2s}\,,
\ene
$$ C=s_1(s+m^2)-(s-m^2)^2\,, \ \lambda_i=\sqrt{(s-s_i)^2-2m^2(s+s_i)+m^4}\,, \ i=1\,, 2\,.$$

The roots of the second factor in the expression B define the region
of variation of $s_1$ and $s_2$ which is shown in the left panel of Fig.~2.

\begin{figure}
\includegraphics[width=0.4\textwidth]{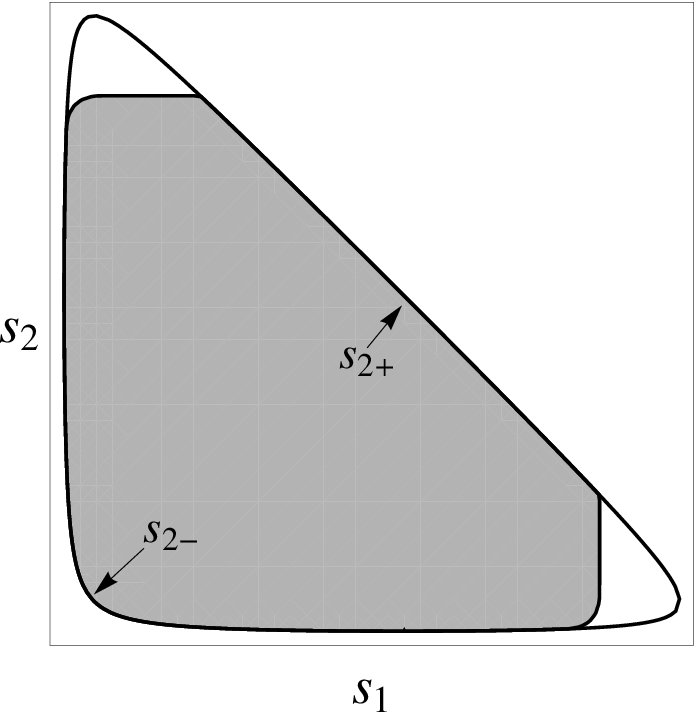}
\hspace{0.4cm}
\includegraphics[width=0.4\textwidth]{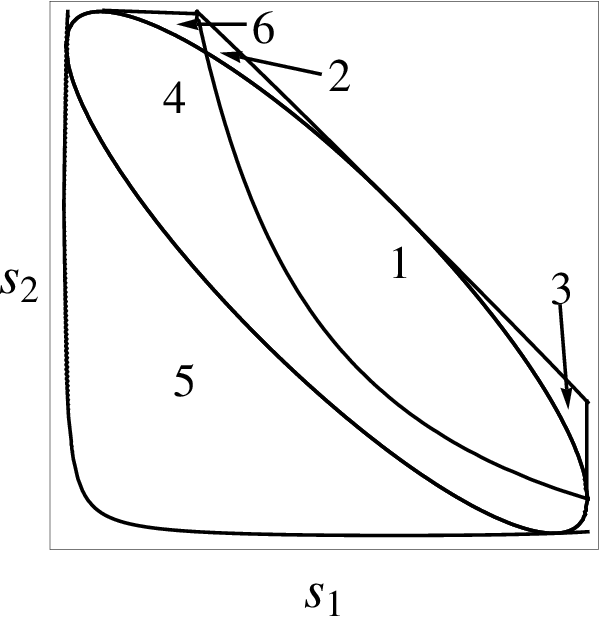}

\caption{The allowed invariant masses $s_1$ and $s_2$ for the
unrestricted (left panel, the total region) and the restricted by
the inequalities (\ref{eq:24}) (left panel, the shaded region) at
$s=10^{-4}$GeV$^2$. The six regions with independent integration
boundaries over the variables $t_2$ and $t_1$ are shown in right
panel.}
\end{figure}

\vspace{0.5cm}

This symmetrical region is restricted by the lines
\bge\label{eq:7}
s_{2-}<s_2<s_{2+}\,, \ \ 4m^2<s_1< (\sqrt{s}-m)^2\,, \ \ s_{2\pm}=\frac{1}{2}\bigg[s-s_1+3m^2\pm \lambda_1\sqrt{1-\frac{4m^2}{s_1}}\,\,\bigg]\,.
\ene

The boundaries of the integration with respect to the variables
$t_2$ and $t_1$, given by Eqs.\,(\ref{eq:5}, \ref{eq:6}), are valid
only for the unrestricted phase space in Fig.\,2 defined by
Eq.\,(\ref{eq:7}).

\subsection{Calculation of the QED cross-section}

For unpolarized particles we must average over the initial particles
polarizations and sum with respect to the final ones . The
differential cross section, in this case, can be written in the form
\bge\label{eq:8}
d\sigma=\frac{1}{4}\,\frac{e^6}{4(k\,p)\,(2\pi)^5}\,\frac{1}{2}\,\sum_{pol}|M|^2\,dR_3\,,
\ene where $M$ is the matrix element of the process (1) which takes
into account the contributions of eight diagrams.  To derive the
distribution over the variables $s_1$ and $s_2$, one has to detect
both final electrons, and additional factor 1/2 in Eq.~(\ref{eq:8})
before $\sum_{pol}|M|^2$ arises due to their identity.

The total matrix element squared of the process can be written as follows
\bge\label{eq:9}
\sum_{pol}|M|^2=|M_b|^2+|\overline{M}_b|^2+|M_c|^2+|\overline{M}_c|^2-2\mathfrak{Re}\big(M_b\overline{M}^*_b\Big)-
\ene
$$-2\mathfrak{Re}\big(M_c\overline{M}^*_c\big)-2\mathfrak{Re}\big(M_b\overline{M}^*_c\big)-2\mathfrak{Re}\big(\overline{M}_bM^*_c\big)+
2\mathfrak{Re}\big(M_bM^*_c\big)+2\mathfrak{Re}\big(\overline{M}_b\overline{M}^*_c\big),
$$ where index $b\,\,(c)$ corresponds to the diagrams of the
Borsellino (Compton-like) type and the bar means the permutation of
the final electrons in the corresponding diagrams.

The double differential cross section with respect to the variables
$s_1$ and $s_2$ (or the $s_1,\,\,s_2-$distribution) reads
\bge\label{eq:10}
\frac{d\,\sigma}{d\,s_1\,d\,s_2}=\frac{\alpha^3}{64\,\pi(s-m^2)^2}\int\int\frac{d\,t_1\,d\,t_2}{\sqrt{-\Delta}}\frac{1}{2}\sum_{pol}|M|^2\,,
\ene where the limits of integration over the variables $t_1$ and
$t_2$ are determined by the choice of the event selection cuts. If
the whole kinematic region is allowed, these limits are defined by
the restrictions (\ref{eq:5}) and (\ref{eq:6}), and the
$s_1,\,\,s_2-$distribution is symmetrical under the permutation
$s_1\rightleftarrows s_2$, provided the identity of the final
electrons is taken into account.

Every matrix element is the contraction of the corresponding current
$j_\mu$ with the photon polarization 4-vector $A^\mu$
\bge\label{eq:11}
M_b=\frac{1}{t_2}A_{\mu}j_b^{\mu}\,, \ \ \overline{M}_b=\frac{1}{t}A_{\mu}\bar{j}_b^{\mu}\,,
\ene
$$M_c=\frac{1}{s_1}A_{\mu}j_c^{\mu}\,, \ \ \overline{M}_c=\frac{1}{s_2}A_{\mu}\bar{j}_c^{\mu}\,, \ t=(p-p_1)^2=2m^2-s-t_1+s_2\,.$$

The currents corresponding to these four types of the diagrams can be
written as
\bge\label{eq:12}
j_b^{\mu}=\bar u(p_2)\gamma_{\lambda}u(p)\bar u(p_1)\widehat{Q}^{\mu\lambda}v(p_3)\,, \ \
\widehat{Q}^{\mu\lambda}=\frac{1}{2d_1}\gamma^{\mu}\hat
k\gamma^{\lambda}-\frac{1}{2d_3}\gamma^{\lambda}\hat
k\gamma^{\mu}+e\,^{\mu}_{(31)}\gamma^{\lambda}\,,
\ene
$$\bar{j}_b^{\mu}=\bar u(p_1)\gamma_{\lambda}u(p)\bar u(p_2)
\widehat{R}^{\mu\lambda}v(p_3)\,, \ \
\widehat{R}^{\mu\lambda}=\frac{1}{2d_2}\gamma^{\mu}\hat
k\gamma^{\lambda}-\frac{1}{2d_3}\gamma^{\lambda}\hat
k\gamma^{\mu}+e\,^{\mu}_{(32)}\gamma^{\lambda}\,, $$
$$j^c{\mu}=\bar u(p_1)\gamma_{\lambda}v(p_3)\bar u(p_2)
\widehat{K}^{\mu\lambda}u(p), \ \
\widehat{K}^{\mu\lambda}=\frac{1}{2d_2}\gamma^{\mu}\hat
k\gamma^{\lambda}+\frac{1}{2d}\gamma^{\lambda}\hat
k\gamma^{\mu}+e\,^{\mu}_{(02)}\gamma^{\lambda}, $$
$$\bar{j}_c^{\mu}=\bar u(p_2)\gamma_{\lambda}v(p_3)\bar u(p_1)
\widehat{L}^{\mu\lambda}u(p), \ \
\widehat{L}^{\mu\lambda}=\frac{1}{2d_1}\gamma^{\mu}\hat
k\gamma^{\lambda}+\frac{1}{2d}\gamma^{\lambda}\hat
k\gamma^{\mu}+e\,^{\mu}_{(01)}\gamma_{\lambda}, $$
where $d=(k\, p)\,, \ d_i=(k\,p_i)$ and
$$e\,^{\mu}_{(0i)}
=\frac{p^{\mu}}{d}-\frac{p^\mu_i}{d_i}, \ \ e\,^{\mu}_{(ij)}
=\frac{p^\mu_i}{d_i}-\frac{p^\mu_j}{d_j}, \ \ i,j=1,2,3. $$

It is easy to verify that every current in the relations
(\ref{eq:12}) satisfies the condition $j^\mu\,k_\mu=0.$ If to define
\bge\label{eq:13}
J^\mu=\frac{j_b^\mu}{t_2}-\frac{\bar{j}_b^\mu}{t}+\frac{j_c^\mu}{s_1}-\frac{\bar{j}_c^\mu}{s_2}\,,
\ene then, for the case of unpolarized particles, we can write
\bge\label{eq:14} \sum_{pol}|M|^2=-g_{\mu\nu}\,J^\mu\,J^{\nu^*}\,.
\ene

The right hand side of Eq.~(\ref{eq:9}) includes three different
structures: four squares of the separate matrix elements, four their
interferences, which enter with negative sign, and two interferences
entering with positive sign. To find the total matrix element
squared, it is enough to calculate only one contribution for every
structure and the other ones can be obtained by means of definite
substitutions. We calculate directly $|M_b|^2\,, \
M_b\,\overline{M}_b^*$ and $M_b\,M_c^*.$ Result reads

 $$ |M_b|^2=\frac{8 }{t_2^2}\bigg\{-4\,m^2 \Big(-\frac{2 d^2}{d_3^2}+\frac{d_3}{d_1}+\frac{d_1}{d_3}\Big) +8 \,(p\,p_3)^2
   \Big[\Big(\frac{1}{d_3}+\frac{1}{d_1}\Big)^2 m^2-\frac{t_2}{d_1\, d_3}\Big]+$$
   $$ +4\,(p\,p_3) \Big[ \frac{t_2^2}{d_1\,d_3}+ t_2 \Big(\frac{2 d}{d_1 d_3}+\frac{1}{d_3}-\frac{1}{d_1}\Big)-
 \frac{4m^2\,d }{d_3}\Big(\frac{1}{d_3}+\frac{1}{d_1}\Big)-m^2\,t_2 \Big(\frac{1}{d_3}+\frac{1}{d_1}\Big)^2\Big]+$$
   $$+t_2^2 \Big[m^2\Big(\frac{1}{d_1}-\frac{1}{d_3}\Big)^2 -\frac{2(d+d_1)}{d_1\,d_3}\Big]+4m^2 t_2
   \Big(\frac{1}{d_3}+\frac{1}{d_1}\Big)\Big[m^2\Big(\frac{1}{d_3}+\frac{1}{d_1}\Big)+\frac{d}{d_3}-2\Big]-$$
\bge\label{eq:15}
   -\frac{t_2^3}{d_1\,d_3}-2\,t_2
   \Big(\frac{d_1}{d_3}+\frac{d_3}{d_1}+\frac{2 d \left(d-d_3\right)}{d_3 d_1}\Big)\bigg\}\,,
\ene

     $$M_b\,\overline{M}_b^*=\frac{8 }{t_2\,t}\bigg\{t_2\Big[\frac{5 d^2-10 d_1 d+d_3^2}{d_1\,d_2}-\frac{m^2(3 d_2^2+4 d_3 d_2+3 d_3^2+2 d d_1)+2m^4(d_2+d_3)}{d_1\,d_2\,d_3}+$$
  $$ +\frac{8 d}{d_2}+\frac{(d_1+d_2){}^2}{d_1\,d_3}\Big]+t\Big[-\frac{2 d^2 \left(d_1-d_3\right)}{d_1
   d_2 d_3}-\frac{2 (d_1+d_3) m^4}{d_1 d_2 d_3}+ t_2\Big(\frac{(d_1+d_2){}^2+6 d\,d_3}{d_1\,d_2\,d_3}-$$
   $$-2 m^2
   \Big(\frac{1}{d_3}+\frac{1}{d_1}\Big)\Big(\frac{1}{d_3}+\frac{1}{d_2}\Big)\Big)-\frac{\left(3 \left(d_1+d_3\right){}^2+2 d d_2-2 d_1
   d_3\right) m^2}{d_1 d_2 d_3}+$$
   $$+\frac{t_2^2}{2} \Big(\frac{2}{d_1 d_3}+\frac{3}{d_1 d_2}+\frac{1}{d_3 d_2}\Big)+\Big(\frac{3 d}{d_2
   d_3}+\frac{1}{d_1}\Big) \left(d-d_3\right)+\frac{4 d \left(d-d_2\right)}{d_1 d_2}\Big]+$$
      $$+4\,m^6 \Big(\frac{1}{d_3}+\frac{1}{d_1}\Big)
   \Big(\frac{1}{d_3}+\frac{1}{d_2}\Big) -\frac{4 \left(\left(d_1+d_2\right){}^2+3 d d_3\right) m^4}{d_1 d_2 d_3}+t_2^2
   \Big[\frac{3 d}{d_1 d_2}-\frac{1}{d_2}+\frac{d}{d_1
   d_3}-$$
      $$-m^2\Big(\frac{1}{d_3}+\frac{1}{d_1}\Big)\Big(\frac{1}{d_3}+\frac{1}{d_2}\Big)\Big] +t^2 \Big[-m^2\Big(\frac{1}{d_3}+\frac{1}{d_1}\Big) \Big(\frac{1}{d_3}+\frac{1}{d_2}\Big)+\frac{t_2(2d_1+d_2+3d_3)}{2\,d_1\,d_2\,d_3}- $$
   $$-\frac{1}{d_1}+\frac{3 d}{d_1 d_2}+\frac{d}{d_2 d_3}\Big]-2m^2 \Big(3
   -\frac{d}{d_2}+\frac{d}{d_3}-\frac{d}{d_1}\Big)+\frac{t_2^3(d_2+d_3)}{2d_1\,d_2\,d_3}+$$
\bge\label{eq:16}
   +\frac{t^3(d_1+d_3)}{2d_1\,d_2\,d_3}+
   \frac{2 d \left(\left(d_1+d_2\right){}^2+2 d d_3\right)}{d_1 d_2}\bigg\}\,,
\ene

    $$M_b\,M_c^*=\frac{1}{t_2\,s_1}\bigg\{\frac{4 \left(4 m^2+t_2\right) \left(d d_1-d_2 d_3\right) t_2^2}{d\,d_1\,d_2\,d_3}+\frac{8 [ d_1(2d^2+dd_1-d_2d_3
    -dd_2)-dd_2d_3 ]t_2^2}{d\,d_1\,d_2\,d_3}+$$
   $$+\frac{8t_2 \left[3 d_1 d^3+\left(2 d_1^2-d d_2\right) \left(d \left(d+d_1\right)+\left(d_1-d\right) d_2\right)+d_2^2 \left(4
   d_1^2+2 d_2 d_1-d_2 d_3\right)\right] }{d\,d_1\,d_2\,d_3}+$$
   $$+\frac{16 m^2t_2 \left(d-d_2\right) \left(-d_1^2+4 d d_1+d_2^2\right)}{d\,d_1\,d_2\,d_3}-\frac{64
   \left(p\,p_3\right){}^3 \left(d_1+d_3\right){}^2}{d\,d_1\,d_2\,d_3}+$$
   $$+\frac{16 \left(d_1+d_3\right) \left[\left(d_1+d_3\right){}^2+2 d d_2+2 d_1
   \left(d_1+d_2\right)\right]}{d_2\,d_3}+$$
   $$+\frac{32 m^2 \left[\left(d_1+d_2\right){}^3+d_1^2 d_3+d \left(d_2^2+\left(3 d-d_3\right) d_3\right)\right]}{d d_2
   d_3}+$$
   $$+\frac{32(p\,p_3)^2}{d\,d_1\,d_2\,d_3} \Big[t_2\big((d+d_1)(d_1+d_3)-2 d_2 d_3\big)+(d-d_2)(4dd_1-2d_1d_2+d^2-d_2^2)\Big]+$$
   $$+\frac{8(p\,p_3)}{d\,d_1\,d_2\,d_3}\Big[t_2^2(d_2d_1+3d_2d_3-dd_3-3dd_1)-4m^2(d-d_2)(d^2+d_2^2+2d_1d_3)-$$
   $$-4m^2t_2(d_1+d_3)^2 -2t_2(2dd_3^2+2d_2d_3^2+7d^2d_1+d^2d_2-3dd_2d_3-7dd_1d_2-d_2^3-d^2d_3)-$$
\bge\label{eq:17}
   -2(d-d_2)\big(4dd_1(d_1+d_2)+(d-d_2)(d+d_1)^2+(d-d_2)(d_1+d_2)^2\big)\Big]\bigg\}\,.
\ene

To obtain the expressions for $|\overline{M}_b|^2$ and
$\overline{M}_b\,\overline{M}_c^*$, we have to perform the
permutation
 $$\widehat{P}_{12}=( p_1\rightleftarrows p_2)$$
 in Eqs.~(\ref{eq:15}) and (\ref{eq:17}), respectively. Such permutation,
evidently, means the substitutions
\bge\label{eq:18}
d_1\rightleftarrows d_2\,, \ t_2\rightleftarrows t\,,  \  s_1\to
s_2\,,
\ene
in these equations, leaving the quantities $d\,, \ d_3$
and $(p\,p_3)$ unchanged.
 Note that the expression $M_b\,\overline{M}_b^*$ is invariant under these substitutions.

The rest of the contributions in Eq.~(\ref{eq:9}) are obtained similarly
by using another substitutions. If we define operators
\bge\label{eq:19}
\widehat{P}_{03}=(p\rightleftarrows-p_3)\,, \ \widehat{P}=(p\rightleftarrows p_1\,,  \ p_2\rightleftarrows-p_3\,,  \ k\to-k)\,,
\ene
then Eq.~(\ref{eq:9}) can be written as
\bge\label{eq:20}
\sum_{pol}|M|^2=\big(1+\widehat{P}_{12}+\widehat{P}_{03}+\widehat{P}_{12}\,\widehat{P}_{03}\big)|M_b|^2
+2\big(1+\widehat{P}_{12}\big)\overline{M}_b\,\overline{M}_c^*
\ene
$$-2\big(1+\widehat{P}_{12}\,\widehat{P}_{03} +\widehat{P}+\widehat{P}_{12}\,\widehat{P}\big)M_b\,\overline{M}_b^*\,.$$

\subsection{Dark photon contribution}

The effective
interaction Lagrangian of the dark photon with the S\,M
 electromagnetic current \cite{H86} can be written as
$$\mathfrak{L}=i\varepsilon\,e\overline{\psi}(x)\gamma^\mu\,\psi(x)\,A'_\mu(x)$$
where $A'_{\mu}$ is the 4$-$potential of the field $\gamma\,'$ and
the small parameter $\varepsilon$ characterizes the coupling
strength relative the electric charge $e.$ In this approach, the
dark photon has to manifest itself as an intermediate state in the
Compton-like Feynman diagrams with the ordinary Breit-Wigner
propagator for the spin-one particle
$$V^{\mu\nu}(q)=\Big(-g^{\mu\nu}+\frac{q^\mu\,q^\nu}{M^2}\Big)P^{BW}(q^2)\,, \ \ P^{BW}(q^2)=\frac{1}{q^2-M^2+iM\,\Gamma}\,,$$
where M ($\Gamma$) is the mass (total decay width) of the dark
photon.

The width of the dark photon decay to the SM lepton pair is
\begin{equation}\label{eq:21}
\Gamma (\gamma\,'\rightarrow
l^+l^-)=\epsilon^2\,\frac{\alpha}{3M^2}(M^2+2m_l^2)\sqrt{M^2-4m_l^2}= \epsilon^2\,\Gamma_0,
\end{equation}
where $m_l$ is the SM lepton mass. In our numerical calculations we restrict ourselves with analysis of the light dark photon signal and
suppose that its mass $M<200\,MeV$. At this condition the decay $A'\to\,\mu^+\,\mu^-$ is closed and, therefore, $m_l$ in Eq.\,(21) is the electron mass.

Due to the contribution of the dark photon, the matrix elements
$M_c$ and $\overline{M}_c$ are modified as
\begin{equation}\label{eq:22}
M_c \to M_c\,R(s_1)\,, \ \ \overline{M}_c \to \overline{M}_c\,R(s_2)\,, \ \ R(s)=1+s\,\varepsilon^2\,P^{BW}(s)\,,
\end{equation}
and this modification leads to the enhancement of the cross section
in two resonance regions: near $s_1\approx M^2$ and $s_2\approx
M^2$. Just in the resonance, the parameter $\varepsilon$ disappears
in the modification factor $R$ because the decay width $\Gamma
(\gamma\,'\rightarrow l^+l^-)$ is proportional to $\varepsilon^2$
$$R(s=M^2)=1-i\frac{M}{\Gamma_0}\,.$$

Taking into account the dark photon contribution, the modified matrix
element squared can be written as
$$\left|M\right|^2=\left|M_b-\overline{M}_b\right|^2+\left|M_c\right|^2\,\left|R(s_1)\right|^2+\left|\overline{M}_c\right|^2\,\left|R(s_2)\right|^2-
2\,\mathfrak{Re}\big[M_c\overline{M}_c\,^*\,R(s_1)\,R^*(s_2)\big]+$$
$$+2\mathfrak{Re}\big\{\big(M_b-\overline{M}_b\big)\big(M_c^*\,R^*(s_1)-\overline{M}_c^*\,R^*(s_2)\big)\big\}\,,$$
where
$$|R(s)|^2=1+\frac{s\epsilon^2}{D(s)}[2(s-M^2)+s\epsilon^2], \ \
D(s)=(s-M^2)^2+M^2\Gamma^2\,,$$
$$\mathfrak{Re}[R(s_1)R^*(s_2)]=1+\epsilon^2\Big\{\frac{s_2}{D(s_2)}(s_2-M^2)+
\frac{s_1}{D(s_1)}(s_1-M^2)+$$
$$+\frac{s_1s_2\epsilon^2}{D(s_1)D(s_2)} [(s_1-M^2)(s_2-M^2)+M^2\Gamma^2]\Big\}\,.$$

\section{Analysis of the dark photon signal}

At the photon energies more than 10\,MeV, the main contribution to
the cross section arises from the Borsellino diagrams due to the events
with small values of $t_2$ and $t$ \cite{BVMP1994}.
In Fig.\,3 we show the ratios $R^c_b$ (the upper row) and
$\overline{R}^c_b$ (the lower row) of the contributions, into
$s_1\,\,,s_2-$distribution caused by the Compton-like diagrams and the
Borsellino ones, defined as
$$
R^c_b=\frac{d\,\sigma_c}{d\,\sigma_b}\,, \ \ d\,\sigma_{c,\,b}=\frac{d\,\sigma}{d\,s_1\,d\,s_2}\Big(\frac{1}{2}\sum_{pol}|M|^2\,\to |M_{c,\,b}|^2\Big)\,,
$$
\begin{equation}\label{eq:23}
\overline{R}^c_b=\frac{d\,\overline{\sigma}_c}{d\,\overline{\sigma}_b}\,,
\ \ d\,\overline{\sigma}_{c,\,b}=
\frac{d\,\sigma}{d\,s_1\,d\,s_2}\Big(\sum_{pol}|M|^2\,\to
|M_{c,\,b}-\overline{M}_{c,\,b}|^2\Big)\,,
\ene
at $s=10^{-4}$\,GeV$^2$, 0.01\,GeV$^2$ and 1\,GeV$^2$ provided the
whole kinematical region is allowed. In this case, the double
differential cross section can be derived analytically by the
integration with respect to the $t_1$ and $t_2$ variables in the
limits defined by the restrictions (\ref{eq:5}) and (\ref{eq:6}).
The quantity $R^c_b$ does not take into account the identity effects
and it is not symmetrical under $s_1\leftrightarrows s_2$, whereas
the quantity $\overline{R}^c_b$ takes into account the identity
effects and it is symmetrical under this permutation.

\begin{figure}
\includegraphics[width=0.32\textwidth]{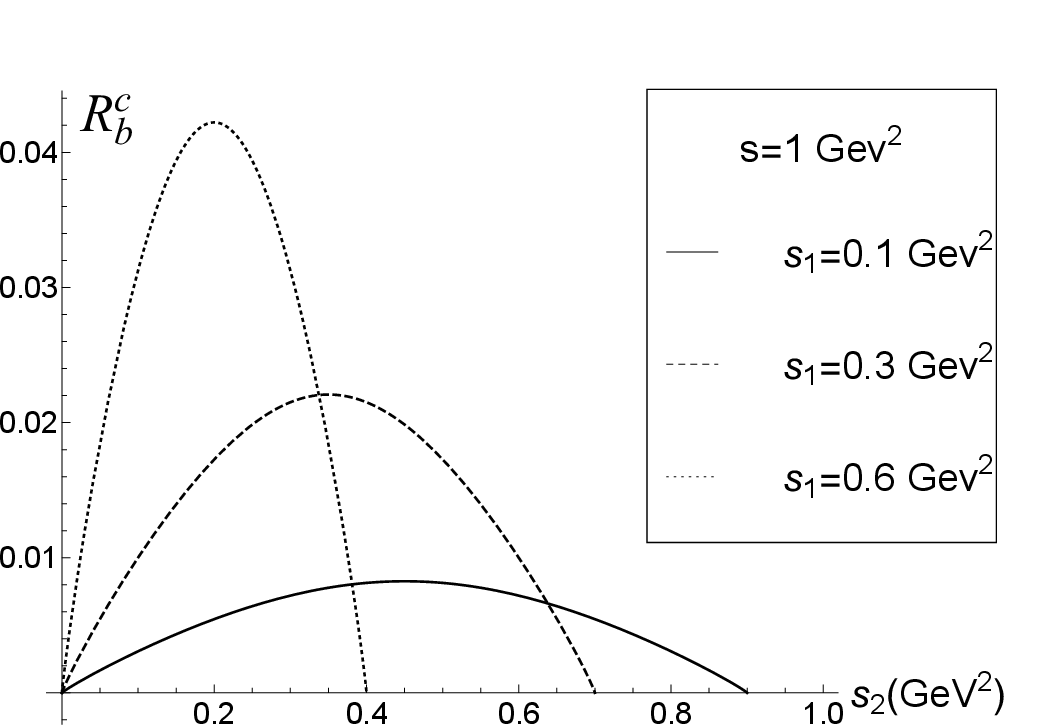}
\includegraphics[width=0.32\textwidth]{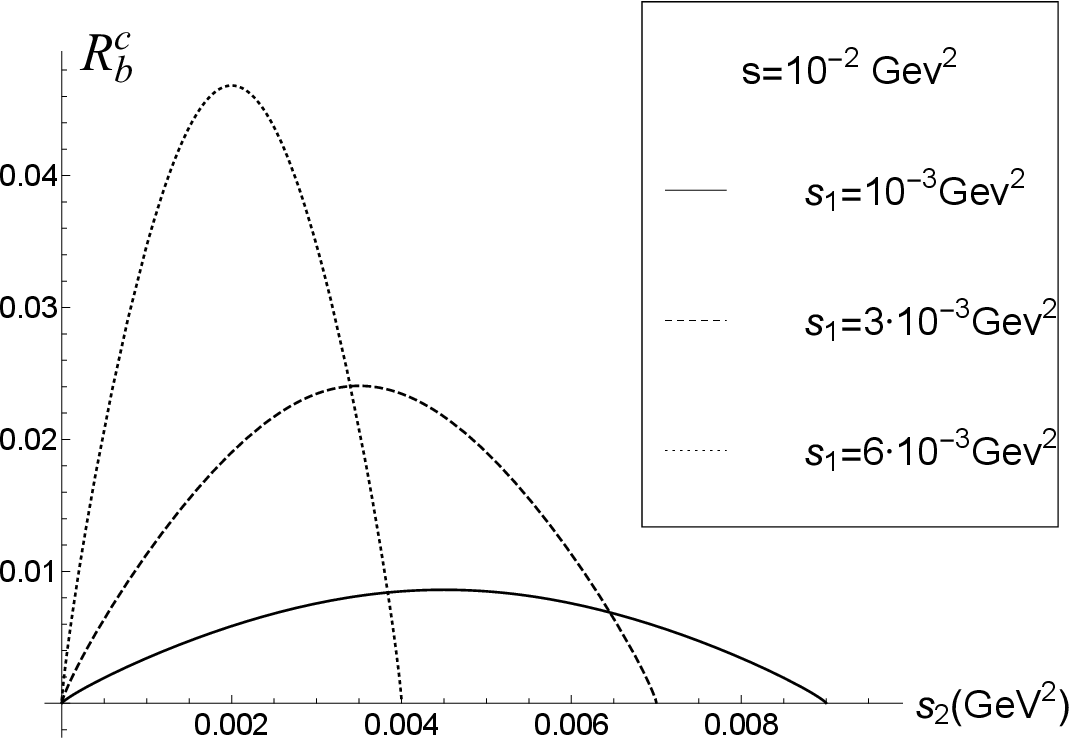}
\includegraphics[width=0.32\textwidth]{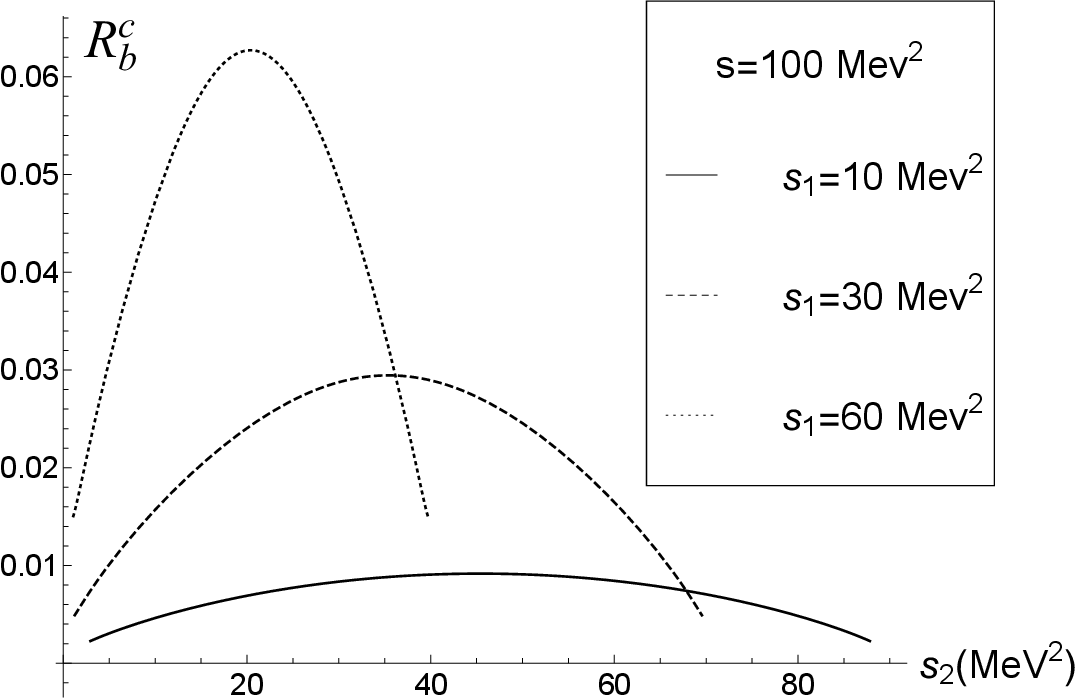}

\vspace{1cm}

\includegraphics[width=0.32\textwidth]{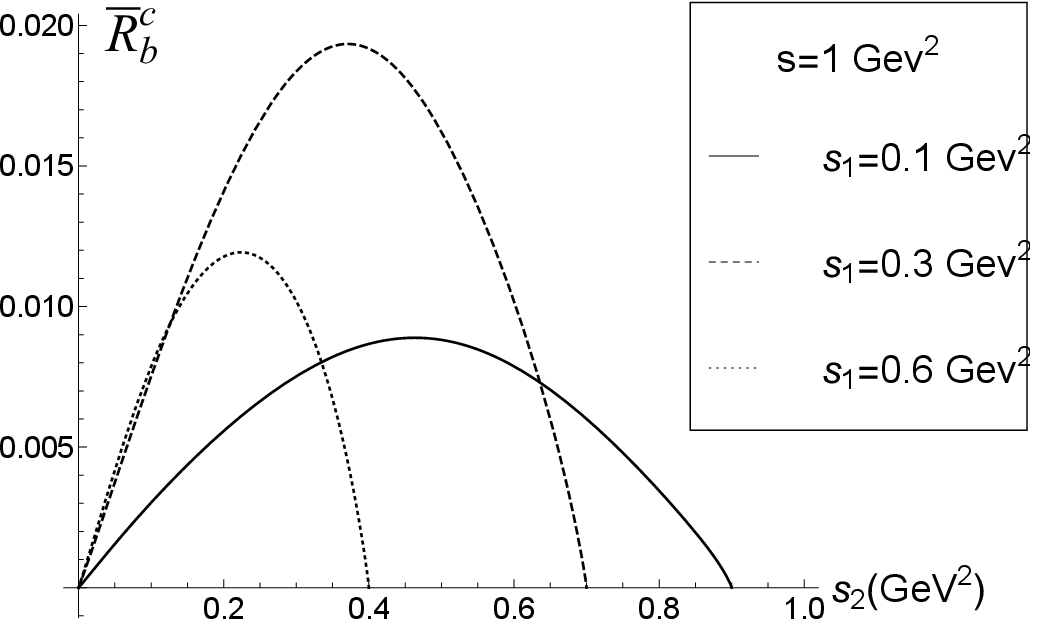}
\includegraphics[width=0.32\textwidth]{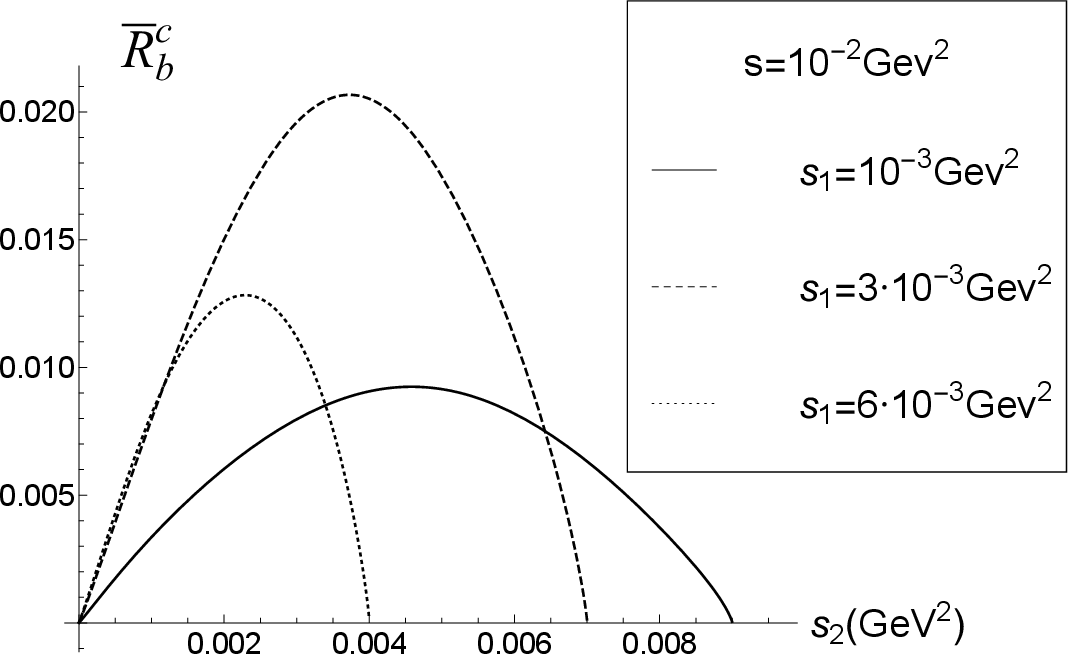}
\includegraphics[width=0.32\textwidth]{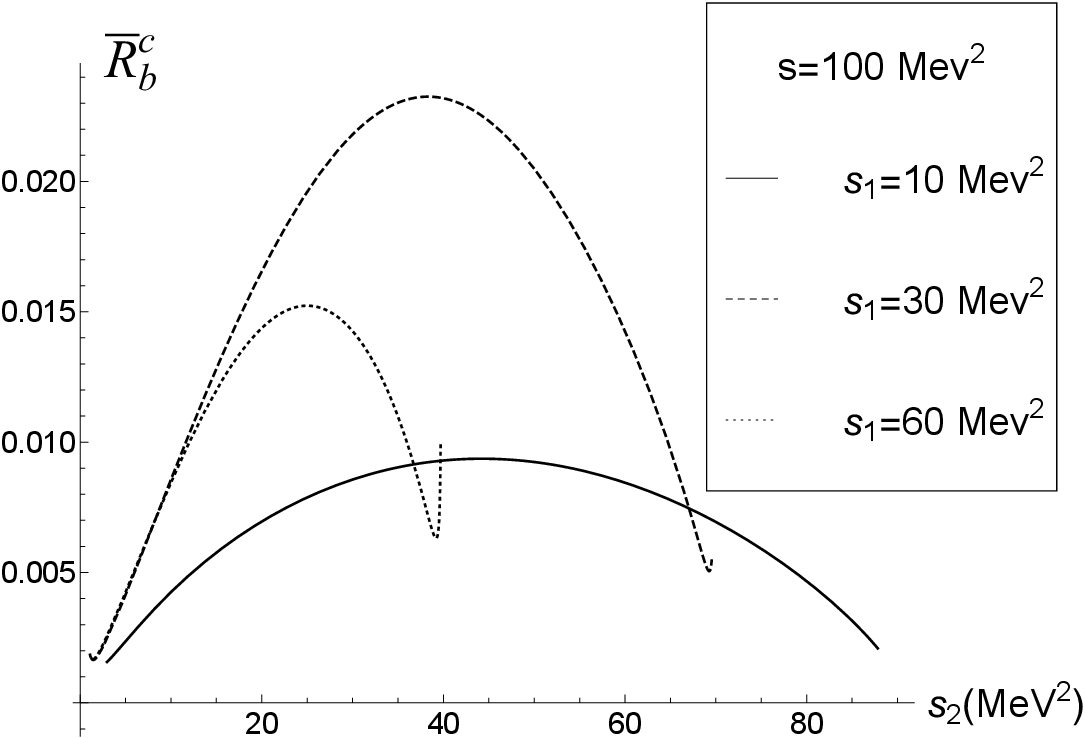}

\caption{The ratio of the double differential distributions over the
invariant masses $s_1$ and $s_2$ for the unrestricted kinematical
region caused by the Compton-like diagrams and the Borsellino ones. The quantity
$R^c_b$  does not take into account the identity effects of the
final electrons and the quantity $\overline{R}^c_b$  takes into
account this effects.}
\end{figure}

In a wide, physically interesting, range of the variables $s_1$ and
$s_2$ the ratio $\overline{R}^c_b$ is rather small (does not exceed
2$\cdot$10$^{-2}$) and it is obvious that, in the case of the
nonrestricted phase space, the Borsellino contribution leads to a very
large QED background to search the not great dark photon signal,
which modifies the Compton-like contribution.

To reduce the Borsellino contribution, we suggest to remove the events,
with small values of the variables $t_2$ and $t$, by the kinematical
cuts
\begin{equation}\label{eq:24}
t_2\,<\eta\,s\,, \ \ t\,<\eta\,s\,, \ \eta<0\,,
\end{equation}
where $\eta$ is a parameter and in our
numerical calculations we use $\eta=-0.2$. It means that, in the rest
frame of the the initial electron, the energy of each final
electrons is more than $0.2\,\omega+1.1\,m\,,$ where $\omega$ is the
photon energy in this frame.

The corresponding symmetrical region of the invariants $s_1$ and
$s_2$ is shown in Fig.\,2 (the shaded region). In this reduced
region
$$4m^2<s_1\,, \, s_2\, <\frac{\eta s(s+m^2)+(s-m^2)\sqrt{\eta s(\eta\,s-4m^2)}}{2\,m^2}\,.$$

In this case, the shaded region in Fig.\,2 is divided into six
independent regions, and each region has its own integration
boundaries over the $t_2$ and $t_1$ variables. The detailed analysis
of the respective kinematics is given in Appendix A. One integration
can be performed analytically and the other one numerically, but the
$s_1\,\,,s_2-$distribution remains symmetrical due to the
symmetrical cuts on $t_2$ and $t$.

The event selection, in accordance with the constraints
(\ref{eq:24}) (with the restricted phase space), decreases
essentially the Borsellino contribution, whereas the Compton-like one is decreased
very little. Their ratio $\widetilde{R}^c_b$, for the restricted
phase space (that is analogue of the ratio $\overline{R}^c_b$), is
shown in Fig.\,4. Due to the final electrons identity,
 there is no possibility to distinguish between the
created electron and the recoil one. Thus, it is necessary to take
into account the effects of the identity.

\begin{figure}
\includegraphics[width=0.32\textwidth]{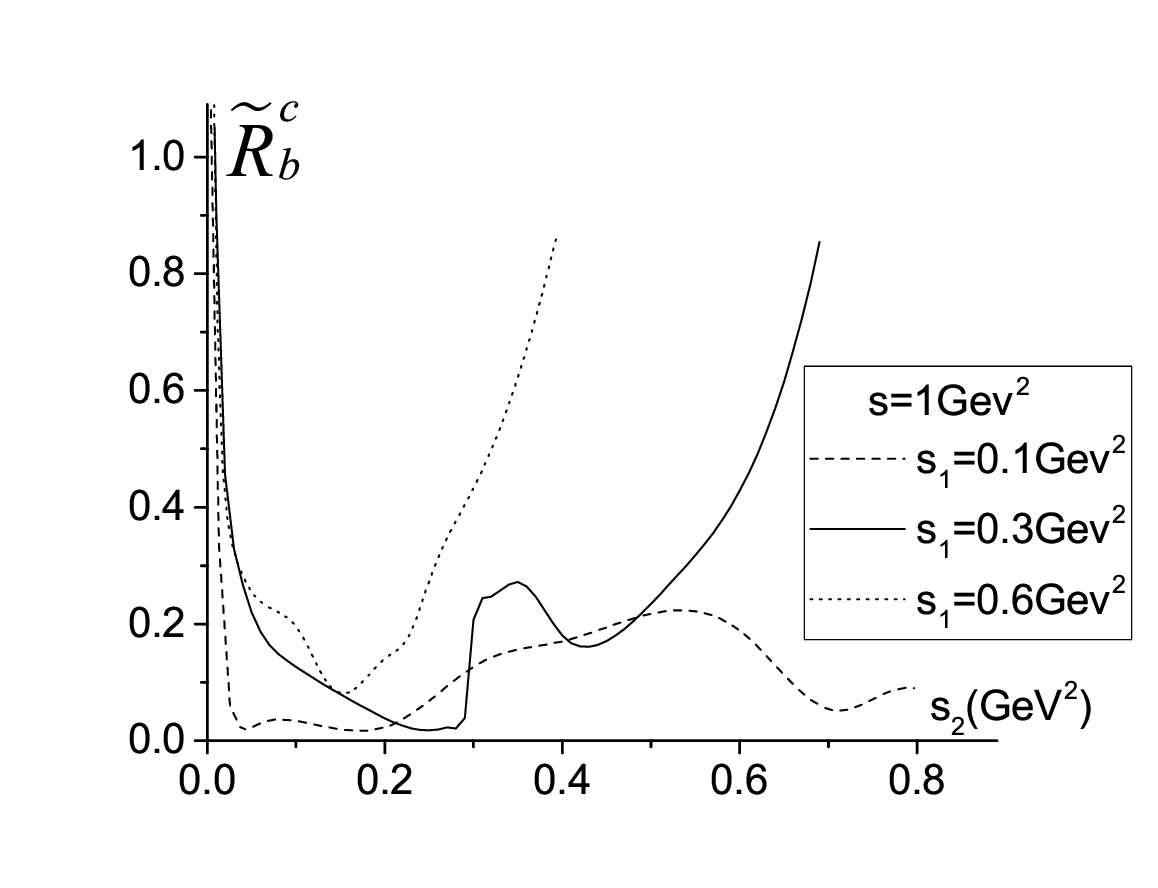}
\includegraphics[width=0.32\textwidth]{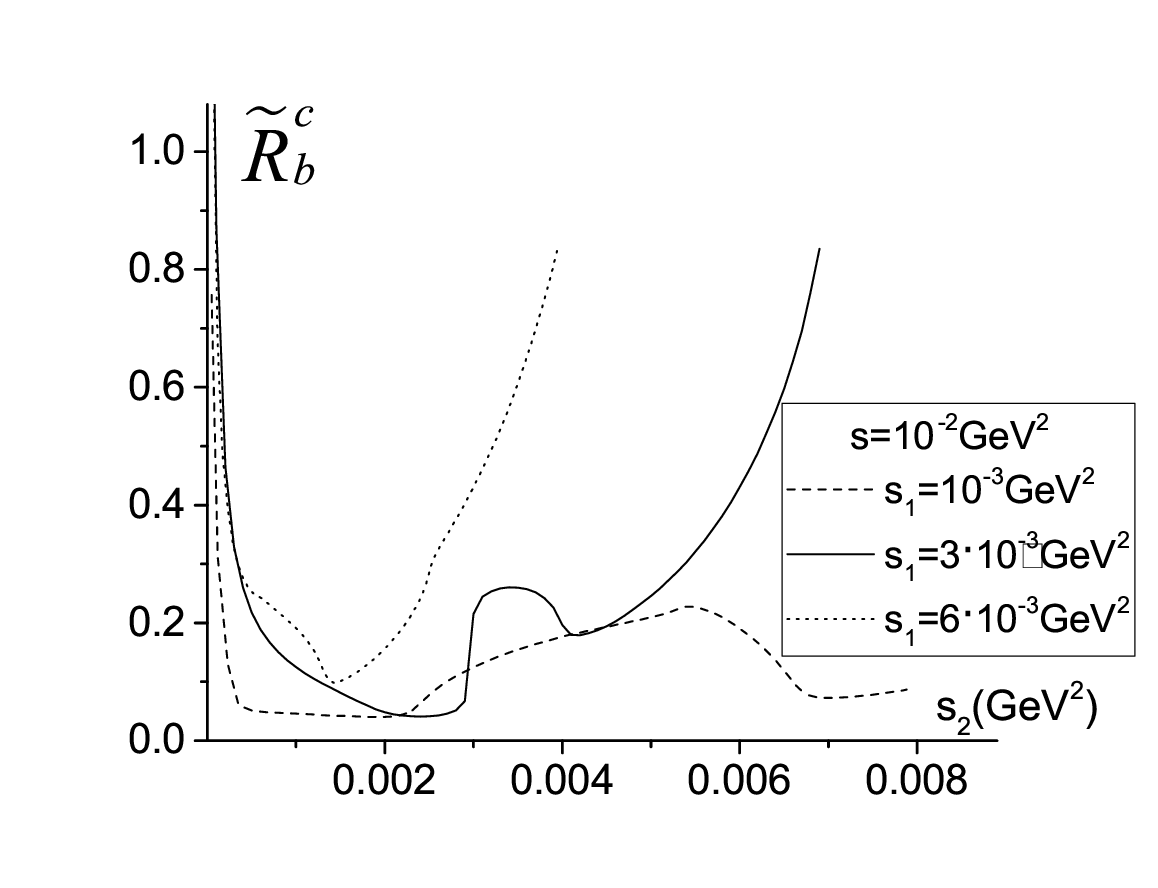}
\includegraphics[width=0.32\textwidth]{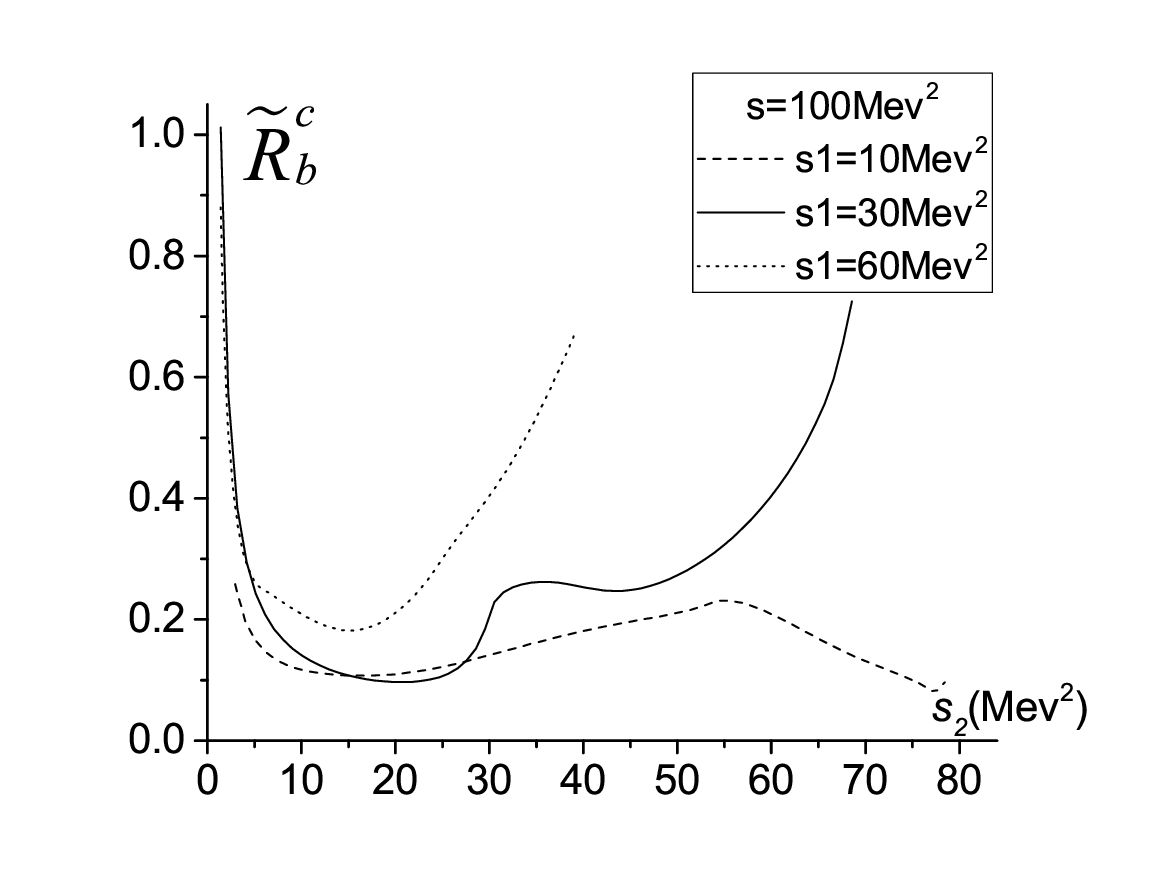}

\caption{The ratio $\widetilde{R}^c_b$ for the restricted
kinematical region, defined by the inequalities (\ref{eq:24}), at
$\eta=-$0.2.}
\end{figure}

In Fig.\,5 we show the double differential cross section (that is
the QED background in search for the dark photon signal in the
process (1)) taking into account all the contributions in the matrix
element squared (\ref{eq:9}) and the constraints (\ref{eq:24}). We
use the dimensionless variables $x_1=s_1/s$ and $x_2=s_2/s$ to see
clearly how quickly this cross section decreases with the increase
of the invariant $s.$ As we noted, the cross section is symmetrical with respect to
the permutation $s_1\leftrightarrows s_2$ and this circumstance
removes the ambiguity of the interpretation of the variables $s_1$
and $s_2$ (due to the final electron identity) in the real
measurement: the event number does not depend what one takes as
$s_1$ or as $s_2.$ To demonstrate this more clearly, we show also
the 3-dimensional plot of the differential cross section at
$s$=10$^{-2}$\,GeV$^2$. In fact, the curves in the middle upper
panel are the intersections of the 3-dimensional plot with the
planes $x_1=0.1, \ x_1=0.3$ and $x_1=0.6.$

\begin{figure}
\includegraphics[width=0.32\textwidth]{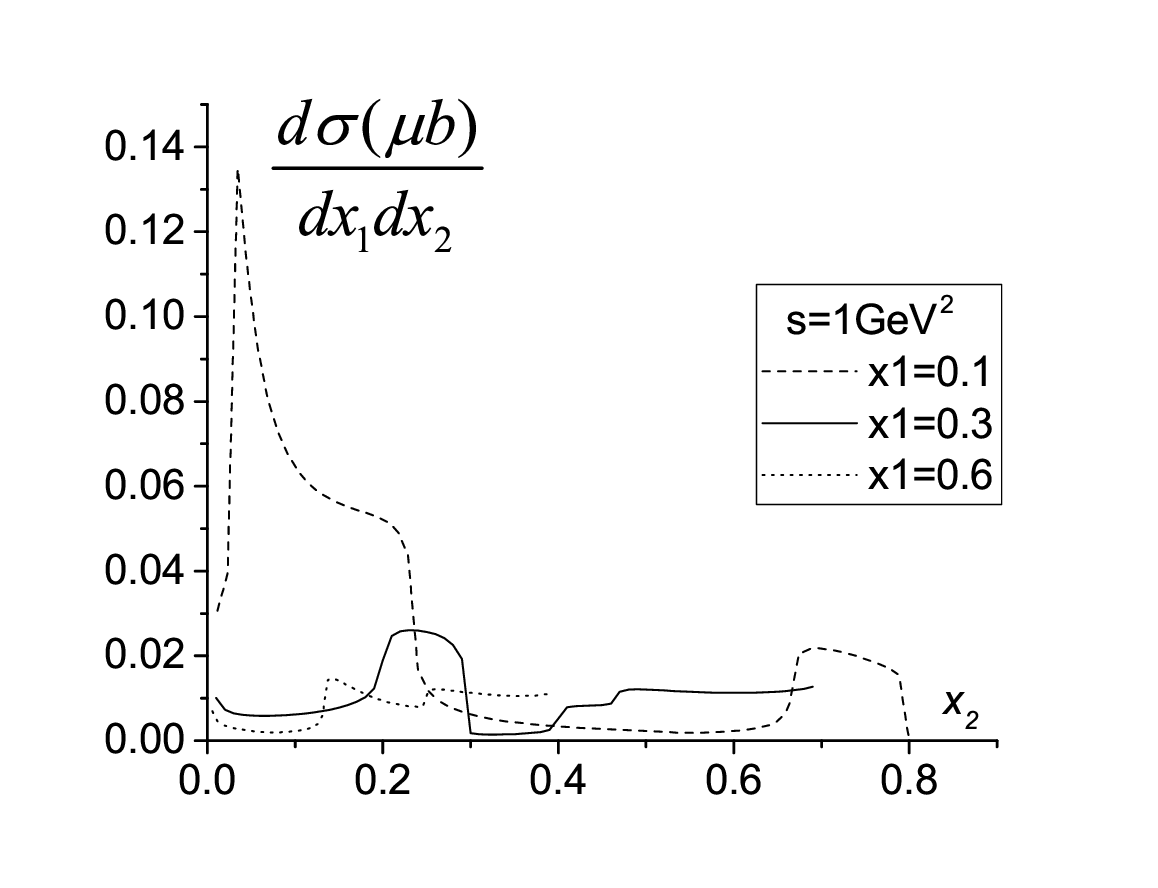}
\includegraphics[width=0.32\textwidth]{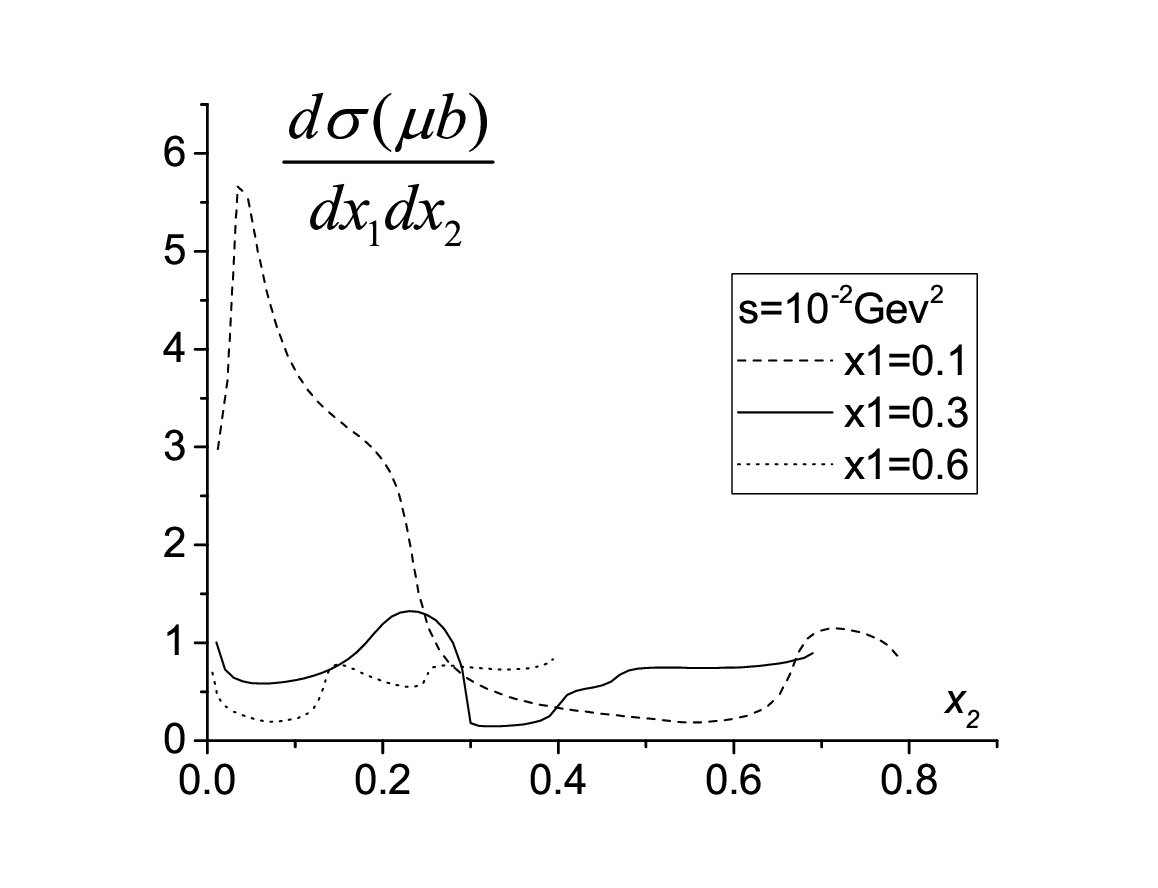}
\includegraphics[width=0.32\textwidth]{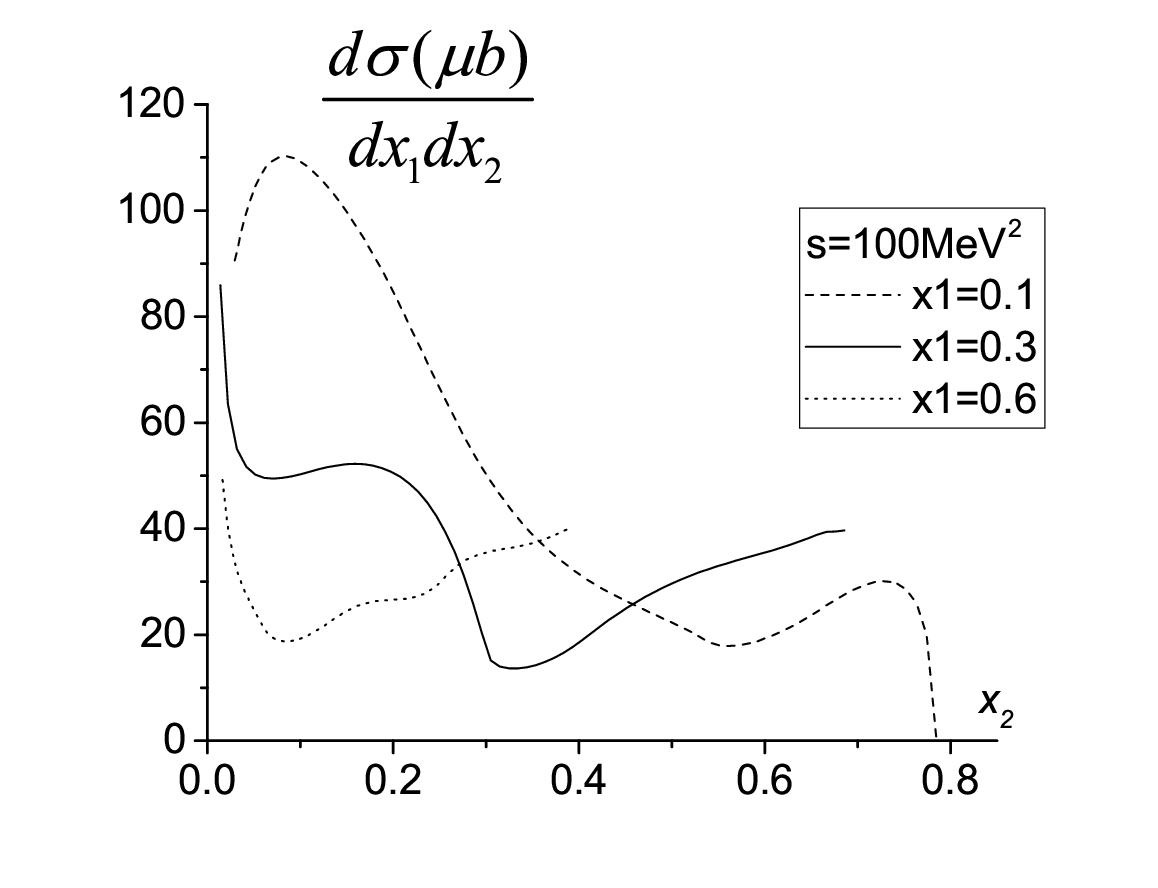}
\includegraphics[width=0.5\textwidth]{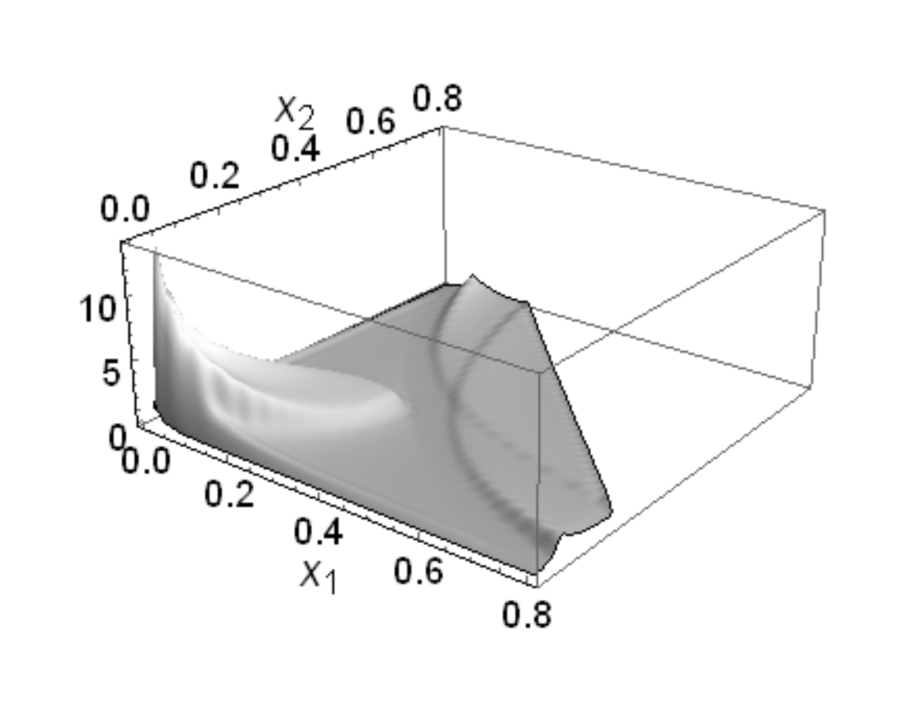}
\caption{The double differential cross section for the restricted
kinematical region, defined by the inequalities (\ref{eq:24}), at
$\eta=-$0.2 as a function of the variable $x_2$ at fixed values of the
variable $x_1.$
The 3-dimensional plot shows the $x_1,\,x_2-$symmetry.}
\end{figure}

Let us estimate the limits for the parameter $\epsilon$ following
Refs. \cite{BEST09,CT16}. We use the definition of the standard deviation
\begin{equation}\label{eq:25}
\sigma =\frac{S}{\sqrt{B}},
\end{equation}
where S(B) is the number of signal (background) events (
$\sigma =2$ corresponds to $\approx $ 95 $\%$ confidence
limit). The event number of any process $i$ is the product of the corresponding
cross section  and the integral luminosity  of the experimental device
$$N_i=d\,\sigma_i\cdot L\cdot T\,,$$
where $L$ is the luminosity, $T$ is the total event accumulation time and all differentials in $d\,\sigma_i$ are dimensionless.

We have also the following relation
\begin{equation}\label{eq:26}
\frac{S}{B}=\frac{d\sigma_{A'}(\varepsilon\,, M^2)}{d\sigma_Q}\,,
\end{equation}
where $d\sigma_{A'}$ is the calculated double differential distribution caused by
the dark photon mechanism
$$d\sigma_{A'}(\varepsilon\,, M^2)=\frac{\varepsilon^2\,s_1[2(s_1-M^2)^2+\varepsilon^2\,s_1]}{D(s_1)}d\,\sigma_c\,,$$
and $d\sigma_Q$ is pure QED contribution with accounting identity of the final electrons. Eqs.~(25) and (26) mean
\begin{equation}\label{eq:27}
\sigma\,d\,\sigma_Q=\sqrt{N}\,d\sigma_{A'}(\varepsilon\,, M^2)\,,
\end{equation}
where $N$ is the number of detected events at particular experimental conditions.
The experimental event selection is: the invariant mass $\sqrt{s_1}$ of the detected $e^+e^--$pair falls in the
energy region
$$M-\delta m/2 < \sqrt{s_1}  < M+\delta m/2\,,$$
where $\delta m$ is the experimental invariant mass resolution, i.e.,
the bin width containing almost all events of possible signal, whereas the invariant mass $s_2$ is fixed.
Then assuming $\Gamma\ll\delta m\ll M$ we can rewrite quantity $D(s_1)$ (see Eq.~(22) and text below) in the approximate form
$$\big[D(s_1)\big]^{-1}=\frac{\pi}{M\,\varepsilon^2\Gamma_0}\delta(s_1-M^2)\,.$$
After integration of the both parts of Eq.\,(27) over the variable $s_1$ within the bin interval $\delta m$ we obtain
\begin{equation}\label{eq:28}
\epsilon^2 =\frac{2\,\sigma}{\pi\sqrt{N}}\frac{\delta
m\Gamma_0}{M^2}\frac{d\sigma_Q (M^2, s_2)}{d\sigma_c (M^2,s_2)}\,.
\end{equation}

The last relation defines the constraints on the parameters $\varepsilon^2\,,\, M^2$ and the detected triplet event number $N$ at given
standard deviation $\sigma$.

In Fig.\,6 we illustrate these constraints at $\sigma=2$ for the energy bin value $1\,MeV$ and the event number $N=10^4.$ For every point in $(\varepsilon^2\,, \ M)$ region (at given values of $s$ and $s_2$) below curves, $\sigma<2$ and above curves,  $\sigma>2$. If the real $A'$ signal corresponds, at least, to three (or more) standard deviations, then the quantities $\varepsilon^2$ (at fixed $M$), when this signal can be recorded, are increased by 1.5 times
(or more) as compared with the corresponding points on the curves in Fig.\,6.

\begin{figure}
\includegraphics[width=0.32\textwidth]{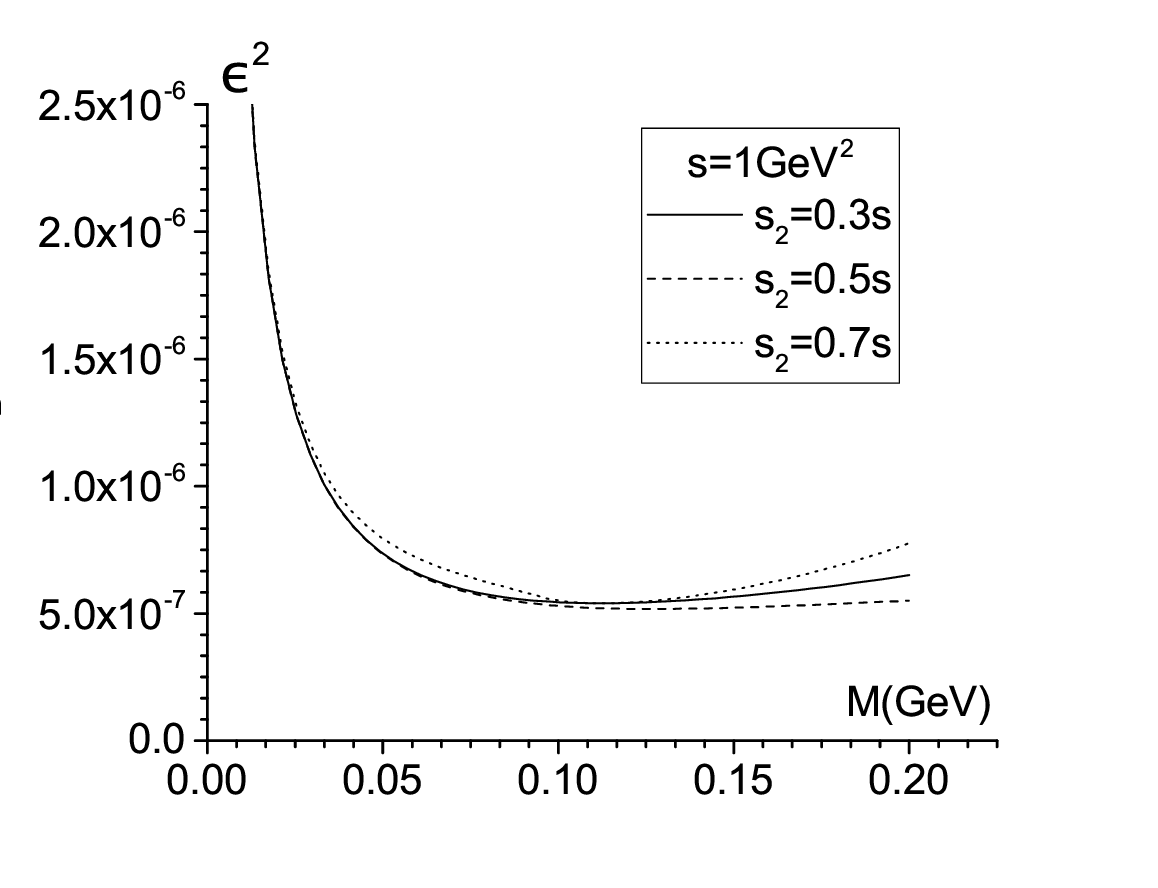}
\includegraphics[width=0.32\textwidth]{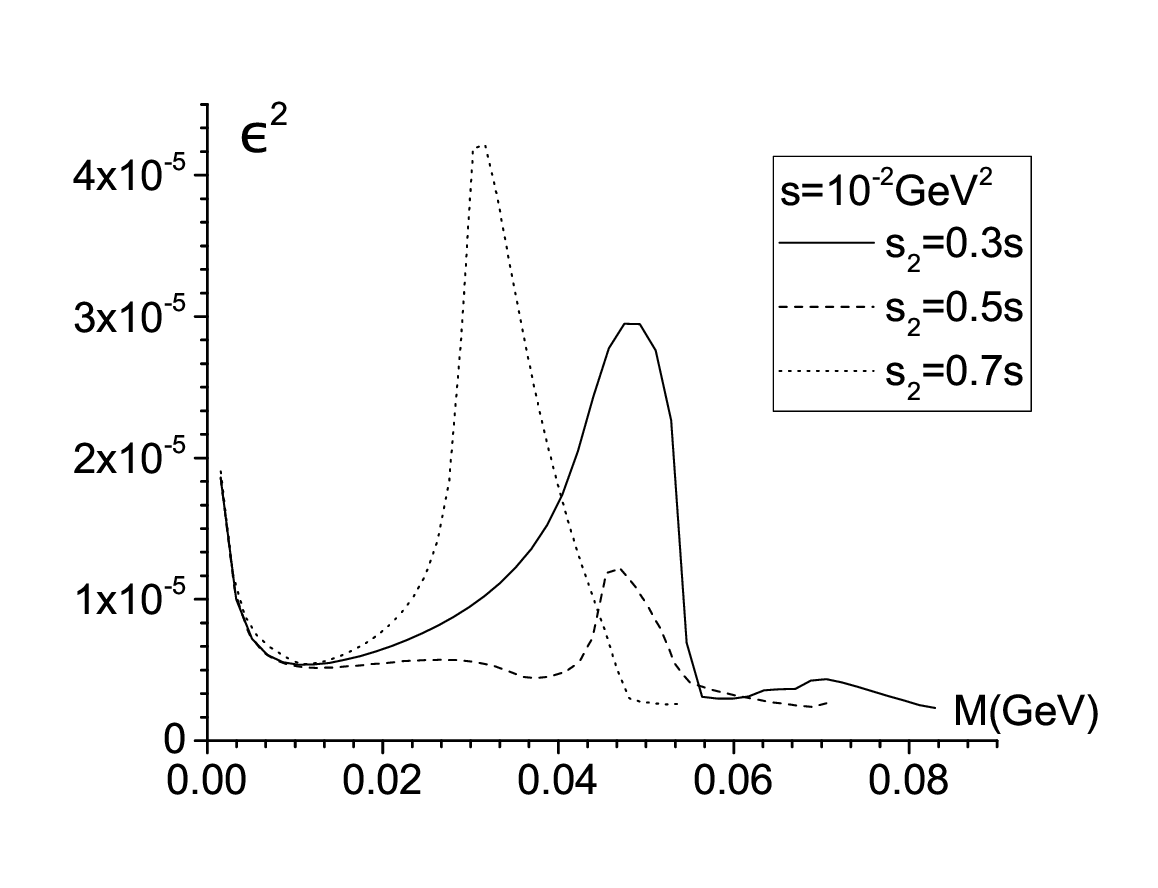}
\includegraphics[width=0.32\textwidth]{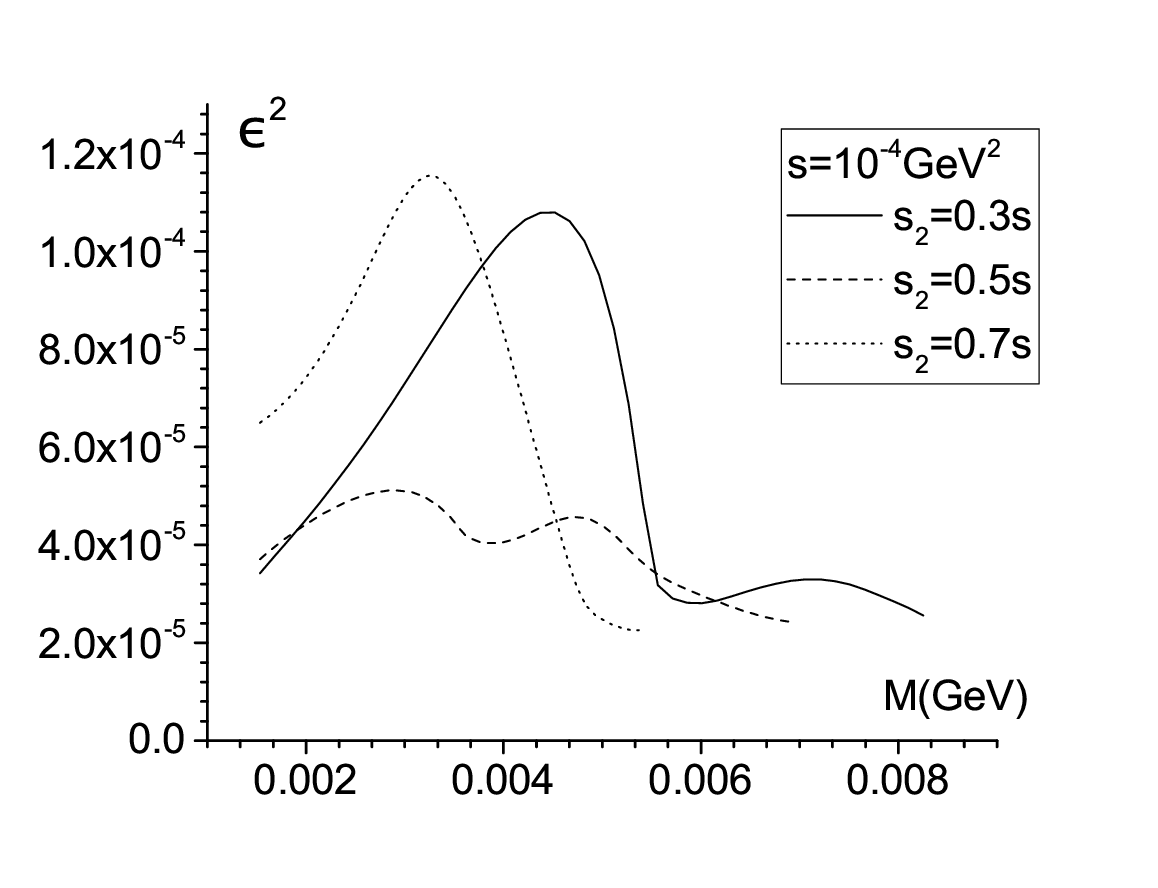}
\caption{The correlation between the $A'$ parameters $\varepsilon^2$ and $M$
at standard deviation $\sigma=2$ and fixed number of recorded triplet $N=10^4$ in the case
of event selection described in text.}
\end{figure}

It is easy to see from Eq.\,(28) that increasing of the energy bin value $\delta m$ decreases the sensitivity of the $A'$ signal detection in the process (1).
The reason is transparent because such experimental device increases the QED background which is, in fact, proportional to $\delta m$ and leaves changeless
the events number caused by the narrow $A'$ resonance.
The dependence of this sensitivity on the
dark photon mass $M$ is not smooth (especially at not great values of $s$) and is defined by interplay of the $M$-dependences of $\Gamma_0,\, d\sigma_Q$ and $d\sigma_c$ entering Eq.\,(28). The account for the kinematical restrictions (24) also increases the sensitivity due to the suppression of the QED background.
Our estimates show that at $s$=1\,GeV$^2$ this suppression is about two orders.

The events number $N$ in the denominator of the right hand side of Eq.\,(28), at described event selection, can be written with a good approximation as
\begin{equation}\label{eq:29}
N=\frac{2\,\delta m\,M}{s}\frac{d\sigma_Q}{d\,x_1\,d\,x_2}\Big(x_1=\frac{M^2}{s}\,, x_2=\frac{s_2}{s}\Big)\,L\cdot\,T \,.
\end{equation}
Using this formula, it is easy to estimate the necessary integral luminosity to accumulate 10$^4$  events. Taking for $d\,\sigma_Q$ values (10$^{-28}$$-$
10$^{-32})$\,cm$^{2}$ (as it follows from the curves in Fig.\,5) and values 10$^{-5}$$-$10$^{-1}$ for $2\,\delta\,m\,M/s$, we obtain the range
$$L\cdot\,T\approx (10\,^{32} - 10\,^{40})\, {\rm cm}^{-2}$$
at considered values of $s$ between 100\,MeV$^2$ and 1\,GeV$^2$.
The largest energies require the largest integral luminosity and vice versa.

The similar estimation were performed by the members of the IRIDE Collaboration (Frascati, Italy) for the
electron-photon collider with the photon range energies (1$-$100)\,MeV and the electron one (100$-$1000)\,MeV, taking into account only the Compton-like
contribution (without final electron identity) \cite{IRIDE}.
Assuming a conservative detector resolution 5\,MeV for the invariant mass of $e^+e^-$-pair
they analysed the sensitivity on $\epsilon^2$ as function of the dark photon mass at different $e^-\,\gamma$ collision energies and in Fig.\,9.8 have
shown their result at integral luminosity 10$^{37}\,{\rm cm}^{-2}$. It indicates an increase of sensitivity for the energy-beam configuration with
photons and electrons of lower energies and this is agreed, at least on the quality level, with our results.

Unfortunately, the search for the light dark photons in the energy region of a few to hundreds MeV in the triplet production,
are not possible at present electron-photon colliders because of the low photon intensities of the machines. But we hope that it will stand  feasible in
a near future.

\section{Conclusion}

In this paper, we have investigated the process of the triplet
photoproduction on a free electron, $\gamma e^-\to e^+e^-e^-,$ which
can be used in the search for a dark photon $A'$. It characterized by
unknown both the mass $M$ and parameter $\epsilon$, describing the coupling
strength relative the electrical charge e, and  can manifest itself in
this reaction being produced as a virtual state with subsequent
decay into $e^+e^-$- pair. The advantage of this
process is that the background to the $A'$ signal is a pure QED process
and it can be calculated with the required precision.

We include the intermediate $A'$ state in two Compton-like diagrams when the virtual dark photon
is time-like, and near the resonance these diagrams can give the observable contribution.
Because near resonance the $A'$ amplitude is mainly imaginary, it practically does not interfere
with pure real QED amplitudes.
Thus, the $A'$ signal is proportional to Compton-like diagram contribution into cross section. As concerns
QED background, in our calculations we take into account all eight Feynmann diagrams.

After trivial azimuthal integration, we performed integration over two squared transferred
momenta $t_1$ and $t_2$, defined by the relations (2), and calculated double differential distribution
over the invariant masses $s_1$ and $s_2$ of two $e^+e^-$ pairs. The boundaries of the variables $t_1$ and $t_2$
for the total phase space of the final particles are obtained from analysis of the Gramian determinant entering
Eq.\,(4). In this case, both integrations are performed analytically (all necessary intermediate results are given in Appendix B) and
the main contribution into cross section, which does not decrease (and even increases logarithmically) when the collision energy grows,
is caused by Borsellino diagrams.

In such situation, the QED background is very large and
to decrease it we restrict the phase space
of the final particles to suppress mainly the Borsellino contribution (see the inequalities (24)). In fact, these inequalities
exclude two regions where the nondecreasing contribution into the cross section, with the growth of the collision energy, is accumulated.
This procedure requires for the detailed study of the kinematics based on the combined analysis of the Gramian determinant and
above inequalities.  The results of this combined analysis are given in Appendix A. In this case, we perform analytically only one integration
and the other one $-$ numerically. The corresponding effect (due to the restriction of the phase space)
can be seen comparing the Compton-like and Borsellino contribution ratios $\bar{B}^c_b$ and $\tilde{B}^c_b$
shown in Fig.\,3 and Fig.\,4, respectively.

The precise analytical results, taking into account all eight diagrams, for the triplet production, given in Appendixes A and B, are new.

As we noted, the considered double differential cross section is symmetrical with respect to the
permutation $s_1 \rightleftarrows s_2$ and this circumstance removes
the ambiguity of the interpretation of the variables $s_1$ and $s_2$
(due to the final electron identity) in the real measurement: the
event number does not depend what one takes as $s_1$ or as $s_2.$  This cross section is shown in Fig.\,5
as a function of the dimensionless variables $x_1=s_1/s$ and $x_2=s_2/s.$

We estimate what value of the strength coupling parameter $\epsilon$, as a function of
the dark photon mass $M$, can be obtained at given number of the measured events $N$ and the value of the standard deviation $\sigma$
assuming some ideal (it may be not realized experimentally but may be used in Monte Carlo simulation) event selection device.
The correlation between $A'$ parameters at $\sigma=2$ and $N=10^4$, at considered rule for the event selection, is shown in Fig.\,6. The curves in
Fig.\,6 indicate the values of  $\epsilon^2$ and $M$ at which the $A'$ signal can manifest itself on the level of two standard deviations at
different collision energies.
Two standard deviations is not enough to interpret the corresponding effect as a manifestation of a new physics, and ordinary one can speak about it on the level three and more.
It easy to recalculate the curves in Fig.\,6 for arbitrary values of $\sigma$ bearing in mind that, in accordance with Eq.\,(28), $\varepsilon^2\sim\sigma.$
As concerns the radiative corrections in the process of the triplet production, as far as we know, they where calculated in the Weizsacker-Williams approximation for the positron spectrum and the total cross section \cite{MO65,VKM74} and never has been considered in suggested experimental setup.  But we think that radiative corrections can not essentially shift the curves in Fig.\,6 because they are no more than a few percent for both $d\,\sigma_Q$ and $d\,\sigma_c.$

The approximate formula (29) allows to estimate (at chosen event selection) the necessary integral luminosity needed to accumulate $N$ triplet events. For $N=10^4$ we received for it $(10^{32}-10^{40}){\rm cm^{-2}}.$

On the quality level our results relative to the sensitivity on $\epsilon^2$ as function of the dark photon mass at different $e^-\,\gamma$ collision energies are agreed with the IRIDE Collaboration estimations, although only Compton-like diagrams has been considered there. Both calculations indicate an increase of sensitivity for the lower $\gamma\,e^-$ collision energies.

\begin{center}
{\Large{\bf{Acknowledgments}}}
\end{center}

This work was partially supported by the Ministry of Education and
Science of Ukraine (Projects No. 0115U000474 and No. 0117U004866).


\section*{Appendix A}
\setcounter{equation}{0}
\def\theequation{A\arabic{equation}}

In this Appendix, we present the integration region with respect to
the variables $t_1$ and $t_2$ taking into account the constraints
(\ref{eq:24}). Firstly, we introduce the dimensionless variables
$t_{1n}$ and $t_{2n}$ in such a way that
$$t_1=\frac{t_{1n}-z_2}{z_1}\,, \ \ t_2=\frac{t_{2n}-z_4}{z_3}\,, $$
$$z_1=-\frac{1}{t_{1-}-t_{1+}}\,, \ z_2=\frac{t_{1-}}{t_{1-}-t_{1+}}\,, \ z_3=-\frac{1}{t_{2-}-t_{2+}}\,, \ z_4=\frac{t_{2-}}{t_{2-}-t_{2+}}\,,$$
where $t_{1\pm}$ and $t_{2\pm}$ are defined in the relations
(\ref{eq:5}) and (\ref{eq:6}), respectively. If the phase space is
unrestricted, we have
$$\int\limits_{t_{2-}}^{t_{2+}}\int\limits_{t_{1-}}^{t_{1+}}d\,t_1\,d\,t_2=\int\limits_0^1\int\limits_0^1 \frac{d\,t_{1n}\,d\,t_{2n}}{z_1\,z_3}$$
for every point $s_1,\,s_2$ from the total region in Fig.\,1 (left panel).

In the case of the restricted phase space, we have six regions of
the variables $s_1,\,s_2$ with well defined boundaries (see Fig.\,1
right panel), and every point $(s_1,\,s_2)$, from each of these
regions, has its own region of integration over the variables $t_1$
and $t_2$. In Fig.\,7, we show all six different $s_1,\,s_2$ regions
and the corresponding variation regions of $t_{1n}$ and $t_{2n}$ at
$s=0.01\,$GeV$^2$ and $\eta=-0.2$.

The $(s_1,\,s_2)$ region {\bf{1}} is defined as
\begin{equation}\label{eq:A1}
B_1(\eta,\,s,\,s_1)<s_2<T_{1+}(\eta,\,s,\,s_1)\,, \ T_{1\pm}=\frac{A_1\,\pm 2\sqrt{C_1\,D_1}}{F_1}\,, \ B_1=\frac{\overline{A}_1}{2\,\overline{F}_1}\,,
\end{equation}
$$A_1=s^2\big[s(1+3\eta+2\eta^2)+s_1(\eta^2-2\eta-1)\big]-m^2\big[s(3+4\eta)+6\eta ss_1-2s_1^2\big]+m^4\big[s(3+\eta)+s_1\big]-m^6\,,$$
$$C_1=\eta s^2(s+\eta s-s_1)+m^2\big[s_1^2-\eta s(2s+s_1)\big]+\eta sm^4\,, \ F_1=(1+\eta)^2s^2-2m^2(s+\eta s+2s_1)+m^4\,,$$
$$D_1=s^2s_1\eta(1+2\eta)+m^2\big[(1+\eta)^2s^2-ss_1(2+7\eta)+s_1^2\big]-2m^4(s+\eta s-s_1)+m^6\,,$$
$${\overline{A}_1}=s^2(s-s_1)\big[\eta(s-s_1)-s_1\big]-sm^2\big[s_1(\eta+2s_1)+\eta(s^2+4ss_1-s_1^2)\big]+m^4\big[s_1(5s+s_1)-\eta s(s+2s_1)\big]+$$
$$+m^6(\eta s-3s_1)+\lambda_1(s-m^2)\big\{s\big[s_1+\eta(s+s_1)\big]-m^2\big[s(2+\eta)+s_1\big]+2m^4\big\}\,,$$
$${\overline{F}_1}=-s^2s_1+m^2(s^2+s_1^2)-m^4(2s+3s_1)+m^6\,.$$

If the point $s_1,\,s_2$ belongs to the region {\bf{1}}, the
corresponding $(t_{1n},\,t_{2n})$ region {\bf{1}} in Fig.\, 7 is
splitted up into two parts
\begin{equation}\label{eq:A2}
\big\{0\,<t_{1n}\,<1,\,\, \ 0\,<t_{2n}\,<X_{1-}\big\}; \ \big\{Z_1\,<\,t_{1n}\,<1,\, \ X_{1-}\,<t_{2n}\,<Y_1,\big\}\,,
\end{equation}
$$Z_1=\frac{s+\eta s-s_2+t_{1-}-2m^2}{t_{1-}-t_{1+}}\,, \  \ Y_1=\frac{-\eta s+t_{2-}}{t_{2-}-t_{2+}}$$
$$X_{1\pm}=\frac{1}{2}-\frac{G_1\,H_1 \pm 4s\,K_1}{2(s-m^2)\lambda_1\,\lambda_2^2}\,,$$
$$G_1=s^2-s_1s_2-s_{12}(s-m^2)-m^4\,, \ H_1=s(s+2\eta s-s_2)-m^2(2s+s_2)+m^4\,,$$
$$K_1=\sqrt{-C_1(s_1\to s_2)\,K}\,, \ K=s_1s_2(s-s_{12})-m^2(s^2-3s_1s_2)+2m^4s -m^6\,, \ s_{12}=s_1+s_2\,.$$

The constraints on the $(s_1,\,s_2)$ region {\bf{2}} in Fig.\,7
reads
\begin{equation}\label{eq:A3}
T_1(\eta,s,s_1)<s_2<s_{2+}\,, \ B_1(\eta,s,s_1)<s_2<s_{2+}\,, \ s_1<\frac{s-m^2}{2}\,,
\end{equation}
and the $(t_{1n},\,t_{2n})$ integration region {\bf{2}} is
\begin{equation}\label{eq:A4}
\big\{0\,<t_{1n}<\,1,\,\, 0\,<t_{2n}<\,X_{1-}\big\}; \ \ \big\{Z_1\,<t_{1n}<\,1,\,\, X_{1-}\,<t_{2n}<\,X_{1+} \big\}\,.
\end{equation}

In the $(s_1,\,s_2)$ region {\bf{ 3}}, the restrictions on the
variable $s_2$ are the same as in the region {\bf{2}} but they apply
at $s_1>(s-m^2)/2.$ The respective $(t_{1n},\,t_{2n})$ integration
region {\bf{3}} is very simple
\begin{equation}\label{eq:A5}
0\,<t_{1n}<\,1,\,\,0\,<t_{2n}\,<Y_1\,.
\end{equation}

In the $(s_1,\,s_2)$ region {\bf{4}} in Fig.\,7, we have
\begin{equation}\label{eq:A6}
T_{1-}<\,\,s_2\,<\,T_{1+}\,,
\end{equation}
provided an additional condition
$$ G_1\sqrt{-C_1(s_1\to s_2)}-H_1\sqrt{K}\,>\,0$$
is satisfied. The $(t_{1n},\,t_{2n})$ integration region {\bf{4}} is
$$Z_1\,<\,t_{1n}\,<\,1\,, \ \ X_{1-}\,<\,t_{2n}\,<\,Y_1\,.$$

The symmetrical $(s_1,\,s_2)$ region {\bf{5}} in Fig.\,7 can be
defined as
\begin{equation}\label{eq:A7}
s_{2-}\,<\,\,s_2\,<\,T_{1-}\,, \ \ s_1\,<\,T_{1-}(s_1\to s_2)\,.
\end{equation}
The $(t_{1n},\,t_{2n})$ integration region {\bf{5}} reads
\begin{equation}\label{eq:A8}
\big\{Z_1\,<\,t_{1n}\,<\,1\,, \ X_{1-}\,<\,t_{2n}\,<\, X_{1+}\big\}\,; \ \ \big\{0\,<\,t_{1n}\,<\,1\,, \ X_{1+}\,<\,t_{2n}\,<\, Y_1\big\}\,.
\end{equation}

At last, the $(s_1,\,s_2)$ region {\bf{6}} in Fig.\,7 is
\begin{equation}\label{eq:A9}
T_{1+}(s_1\to s_2)\,<\,s_1\,<B_2(\eta,s,s_2))\,, \ s_2\,>\,T_{1-}\,, \ B_2=\frac{A_2+C_2\sqrt{K_2}}{2\,F_2}\,,
\end{equation}
$$A_2=s^2\big[2\eta^2s^2+(s-s_2)(2\eta s-s_2)\big]+s^2m^2\big[2\eta^2s-s_2-2\eta(s+2s_2)\big]+m^4\big[s_2(5s+3s_2)-2\eta s(s+s_2)\big]+$$
$$+m^6(2\eta s-3s_2)\,, \ \
C_2=s(s+2\eta s-s_2)-m^2(2s+s_2)+m^4\,,$$
$$K_2=s^2s_2^2+2sm^2\big[2\eta s^2+2\eta s(s-s_2)-s_2(2s+s_2)+m^4\big[s_2(8s+5s_2)-4\eta s(2s+s_2)\big]+4m^6(\eta s-s_2)\big]\,,$$
$$F_2=(\eta s-s_2)\big[(1+\eta)s^2-m^2(2s+s_2)+m^4\big]\,.$$
The corresponding $(t_{1n},\,t_{2n})$ integration region {\bf 6}
\begin{equation}\label{eq:A10}
Z_1\,<\,t_{1n}\,<\,1\,, \ \ X_{1-}\,<\,t_{2n}\,<\,X_{1+}\,.
\end{equation}

\begin{figure}
\includegraphics[width=0.3\textwidth]{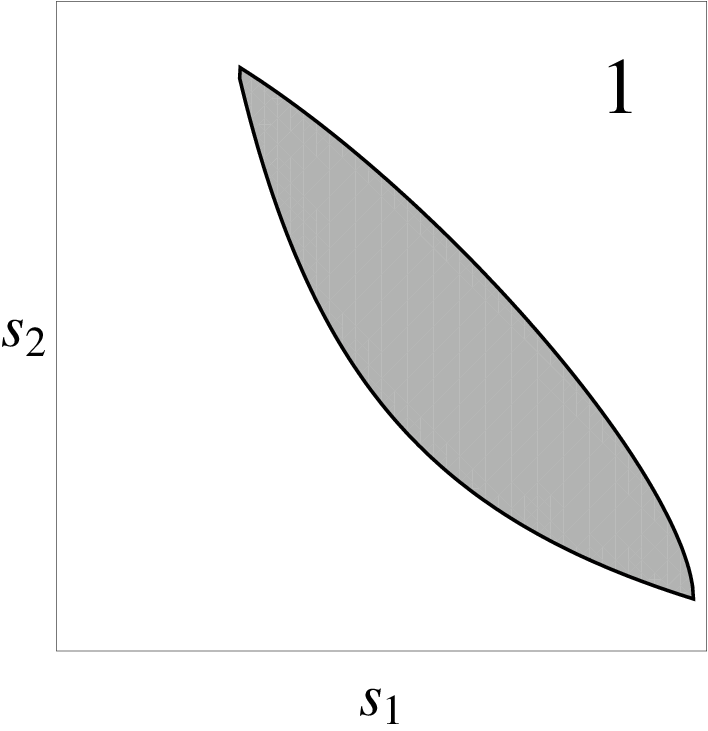}
\hspace{0.2cm}
\includegraphics[width=0.3\textwidth]{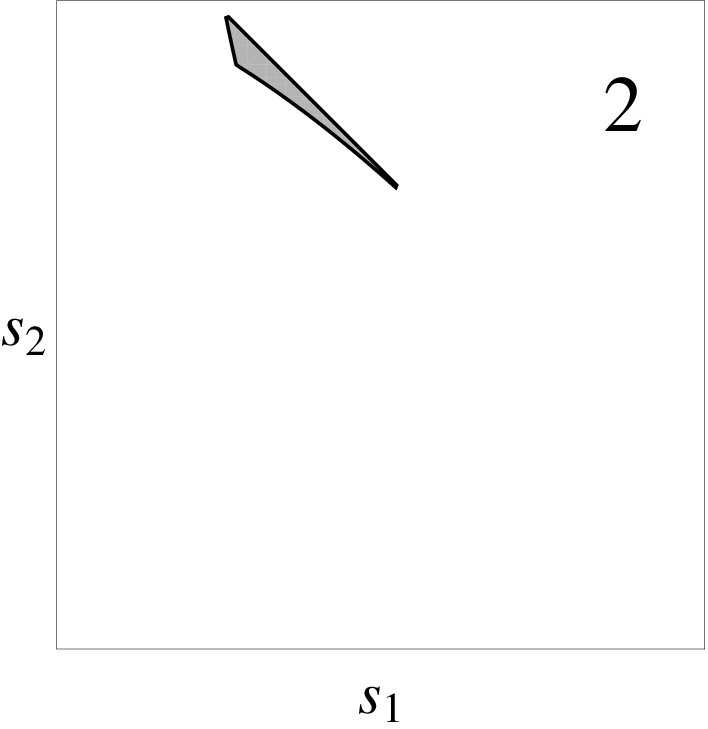}
\hspace{0.2cm}
\includegraphics[width=0.3\textwidth]{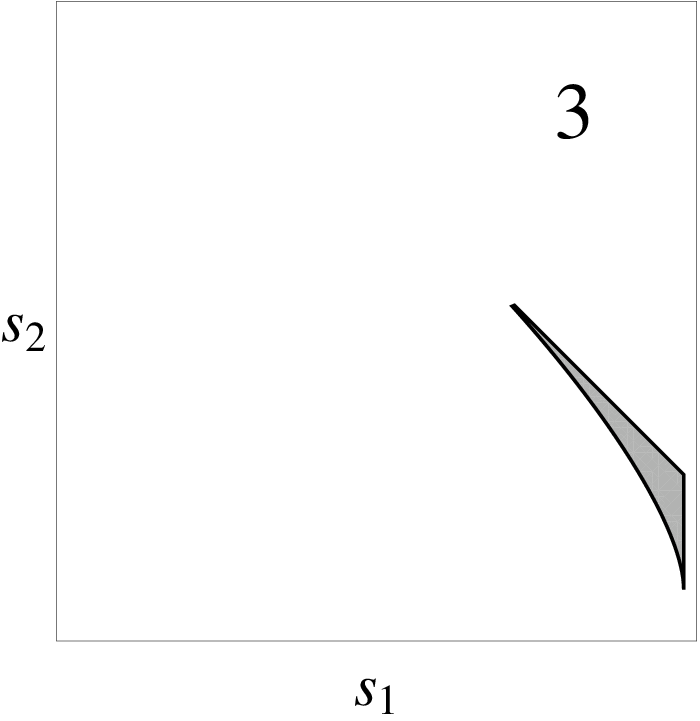}

\vspace{0.15cm}

\includegraphics[width=0.3\textwidth]{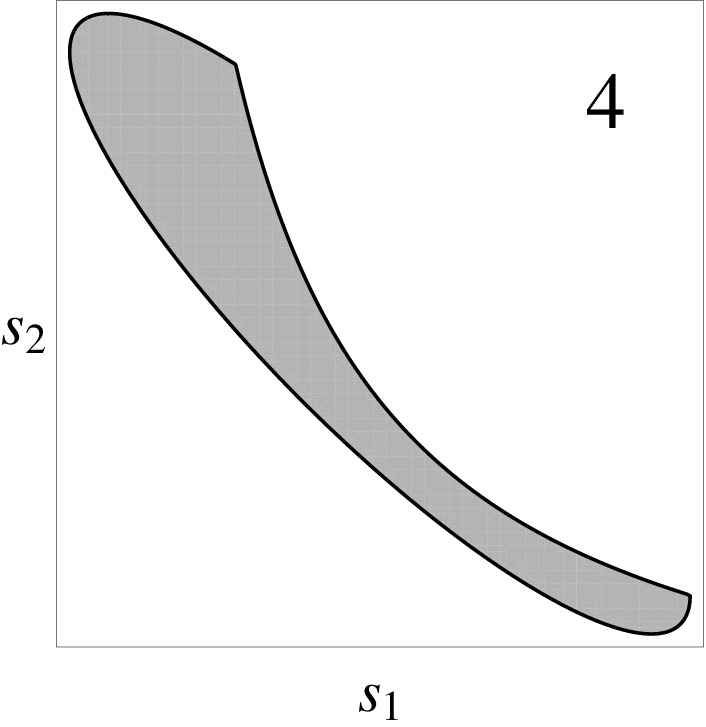}
\hspace{0.2cm}
\includegraphics[width=0.3\textwidth]{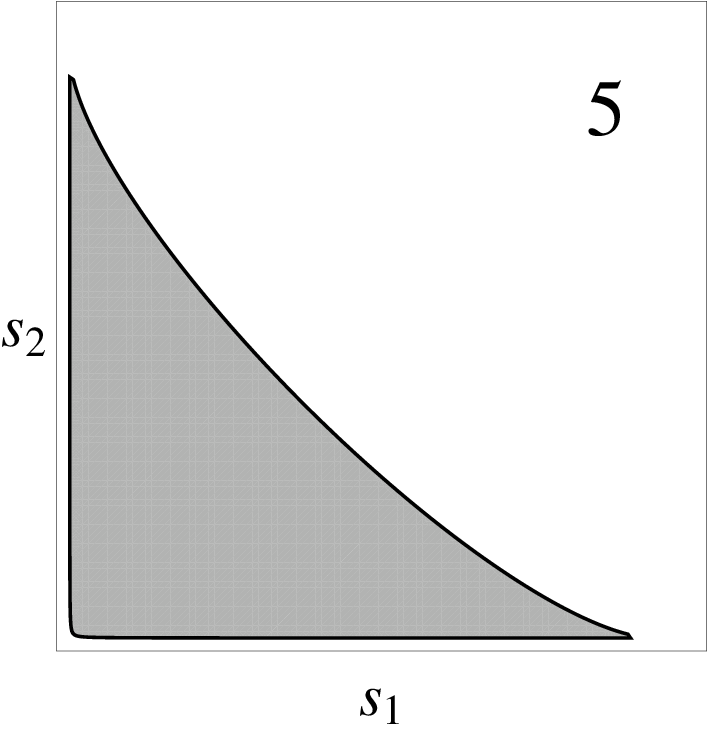}
\hspace{0.2cm}
\includegraphics[width=0.3\textwidth]{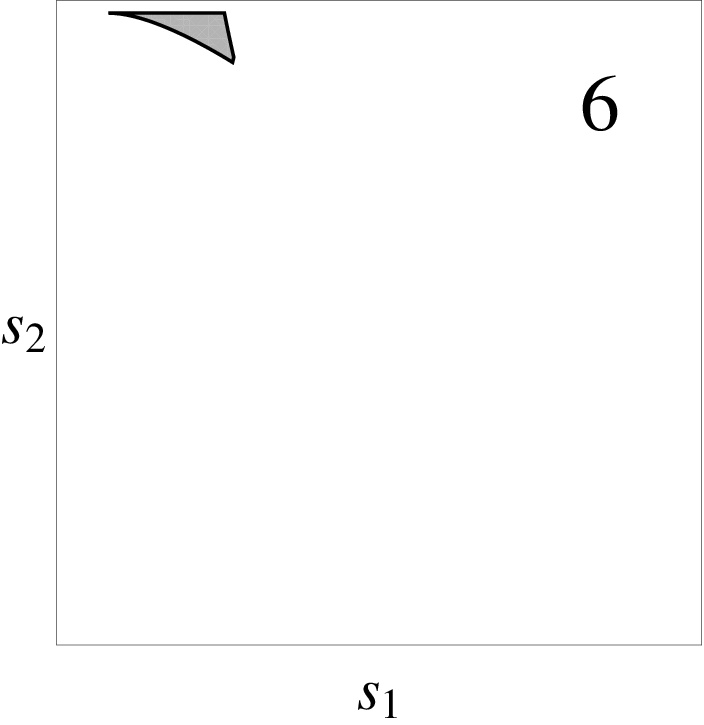}

\vspace{0.15cm}

\includegraphics[width=0.31\textwidth]{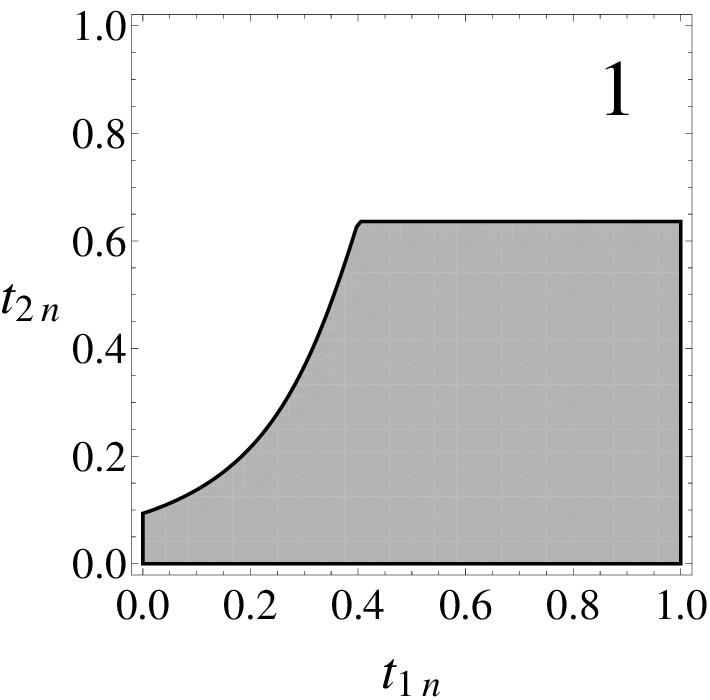}
\includegraphics[width=0.31\textwidth]{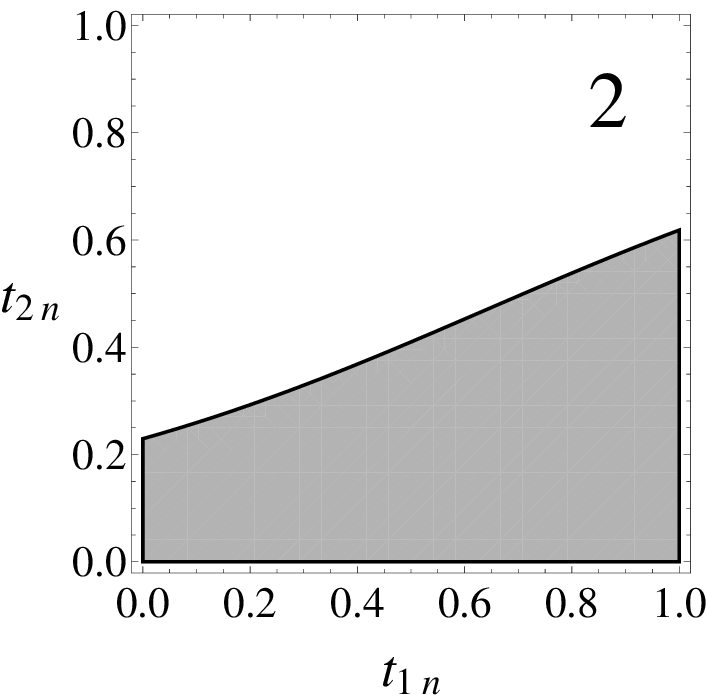}
\includegraphics[width=0.31\textwidth]{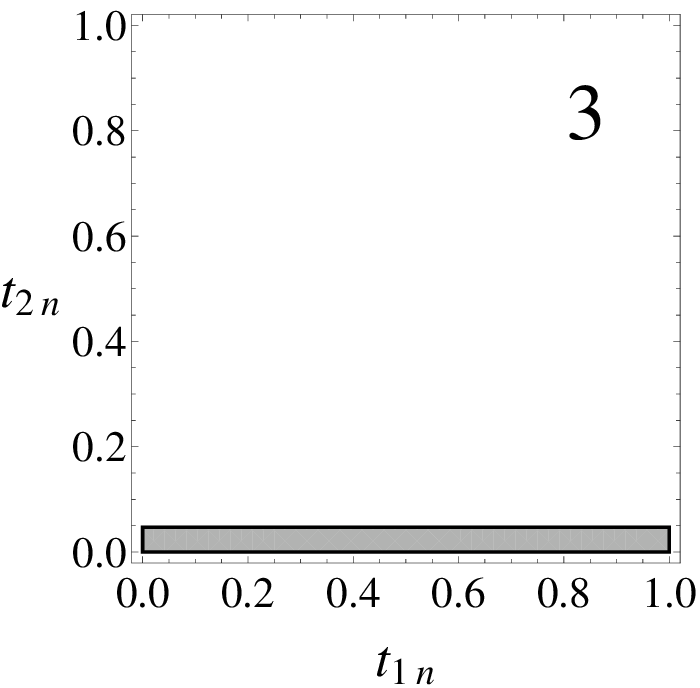}

\vspace{0.15cm}

\includegraphics[width=0.31\textwidth]{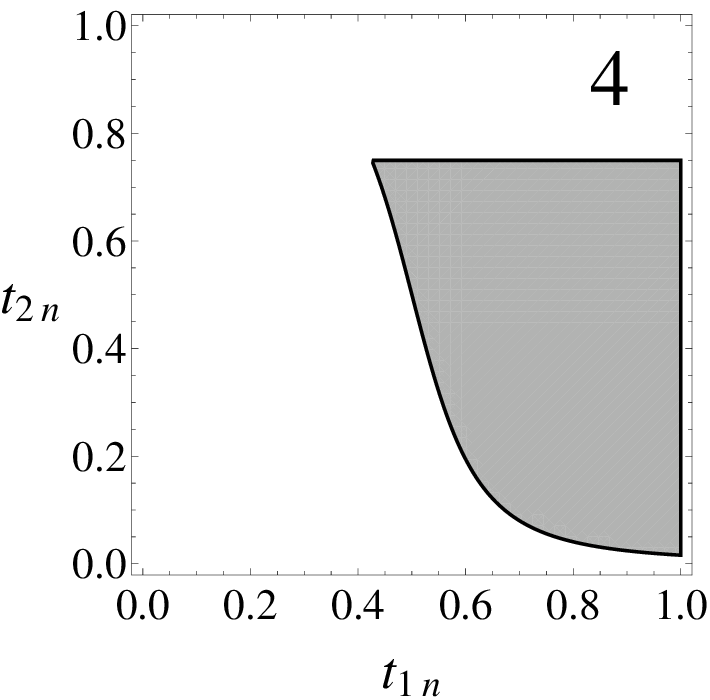}
\includegraphics[width=0.31\textwidth]{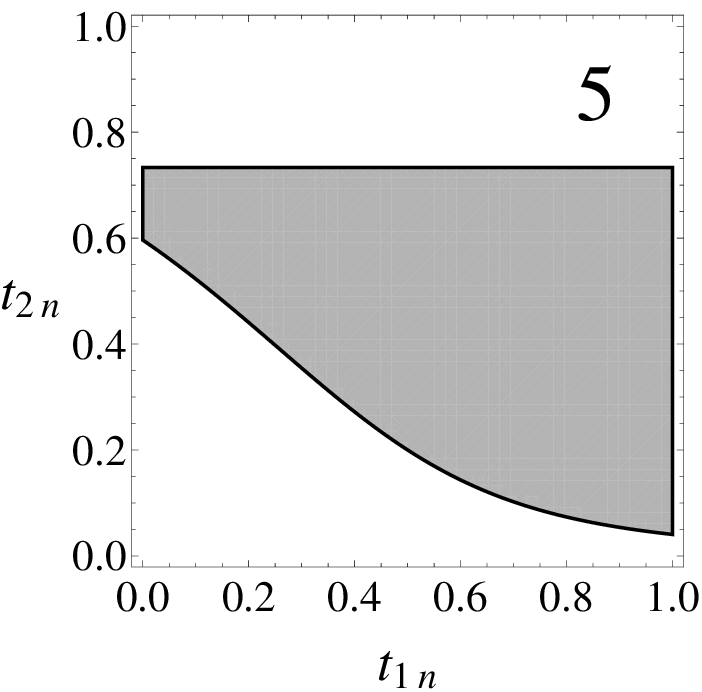}
\includegraphics[width=0.31\textwidth]{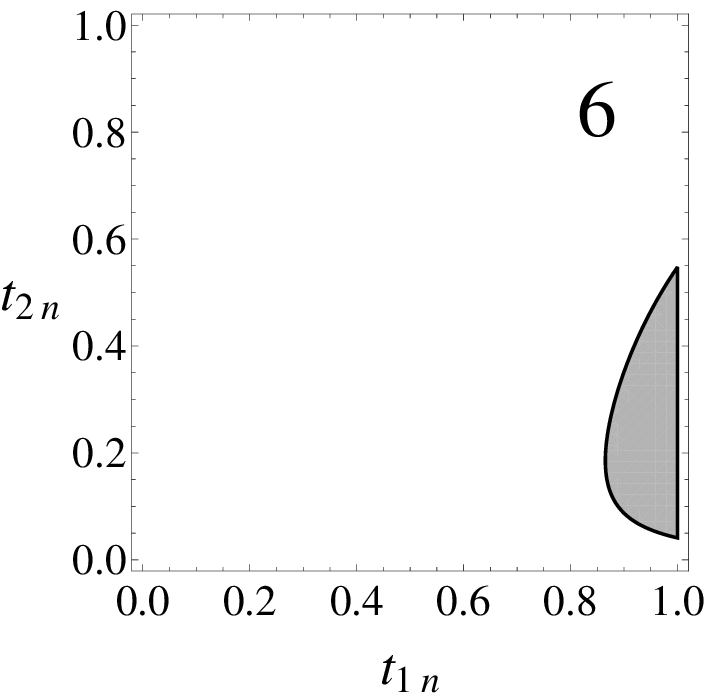}

\caption{The six kinematical regions, defined by the inequalities
(\ref{eq:24}), at $\eta=-$0.2 defined by the inequalities in Appendix A.}
\end{figure}

\section*{Appendix B}
\setcounter{equation}{0}
\def\theequation{B\arabic{equation}}

Here, we write the analytical results for the
contributions of different diagrams and their interferences into the
matrix element squared, as defined by the Eq.\,(\ref{eq:9}),
integrated over the variables $t_1$ and $t_2$ with the weight
$$\frac{\pi}{64\,(s-m^2)\,\sqrt{-\Delta}}\,.$$ We introduce the short notation
$$\frac{\pi}{64\,(s-m^2)}\int\limits_{t_{1-}}^{t_{1+}}d\,t_1 \int\limits_{t_{2-}}^{t_{2+}}d\,t_2\frac{W}{\sqrt{-\Delta}} \equiv \overbrace{W}\,.$$
Then we have
\begin{equation}\label{eq:B1}
\overbrace{|M_c|^2}=\frac{\pi^2}{2s^2s_1^2(s-m^2)^2\lambda_1^4}\Big[A_c+\frac{4s^2\,B_c\,L_1}{\lambda_1}\Big]\,, \ \ L_1=\ln\frac{m^2+s-s_1+\lambda_1}{2m\sqrt{s}}\,,
\end{equation}
$$A_c=s^2s_1 \left(s-s_1\right)\bigl[s^4+2\, s^3
   \left(s_1-s_2\right)+2\,s^2 \left(-7 s_1^2+s_2 s_1+s_2^2\right) +2\, s\, s_1^2 \left(9 s_1+7 s_2\right)-$$
 $$  -7 s_1^2 \left(s_1^2+2 s_2 s_1+2 s_2^2\right)\bigr]+$$
$$+s\,m^2\bigl[2 s^6+17 s^5s_1 -2 s^4s_1 \left(41 s_1+16 s_2\right) +2 s^3s_1 \left(32 s_1^2-42 s_2 s_1+13 s_2^2\right) +$$
   $$+2 s^2s_1^2 \left(19 s_1^2+142 s_2
   s_1+58 s_2^2\right) -s\,s_1^3 \left(41 s_1^2+172 s_2 s_1+126 s_2^2\right) +2 s_1^4 \left(s_1^2+2 s_2 s_1+2 s_2^2\right)\bigr]+$$
$$+m^4\bigl[22 s^6-87 s^5s_1 +s^4s_1 \left(319 s_1-18 s_2\right) -4 s^3s_1
   \left(74 s_1^2+140 s_2 s_1+15 s_2^2\right) +$$
  $$ +2 s^2s_1^2 \left(69 s_1^2+262 s_2 s_1+76 s_2^2\right) +s\,s_1^3 \left(-3 s_1^2-4 s_2 s_1+6 s_2^2\right) -s_1^4 \left(s_1^2+2 s_2 s_1+2 s_2^2\right)\bigr]+ $$
$$+m^6\bigl[-110 s^5+153 s^4s_1 +36 s^3s_1 \left(17 s_1+4 s_2\right) -
4 s^2s_1 \left(74 s_1^2+104 s_2 s_1-9 s_2^2\right) -$$ $$
   -2\,s\,s_1^2 \left(3 s_1^2+14 s_2 s_1+2 s_2^2\right)+ s_1^3 \left(5 s_1^2+12 s_2 s_1+6 s_2^2\right)\bigr]+$$
$$+m^8\bigl[182 s^4-117 s^3s_1 +
   s^2s_1 \left(191 s_1-102 s_2\right) +6 s\,s_1 \left(4 s_1^2+2 s_2 s_1-s_2^2\right) -2
   s_1^2 \left(5 s_1^2+12 s_2 s_1+3 s_2^2\right)\bigr]+$$
$$+m^{10}\bigl[-138 s^3+27 s^2s_1 -2s\,s_1 \left(9 s_1-8 s_2\right)
   +2 s_1 \left(4 s_1^2+10 s_2 s_1+s_2^2\right)\bigr] +$$
$$+m^{12}\left[s\left(50 s+11 s_1\right)+s_1 \left(s_1-6 s_2\right)\right]-5m^{14}(2s+s_1)+2m^{16}\,,$$

$$B_c=\left(s-s_1\right){}^2 s_1 \left(s^2-2 s_1 s+2 s_1^2\right)\big[s^2-2 s \left(s_1+s_2\right)+s_1^2+2 s_2^2+2 s_1 s_2\big]+$$
$$+2\,m^2\bigl[s^6-9 s_1 s^5+s^4s_1 \left(23 s_1+5 s_2\right) -2\,s^3 s_1 \left(11 s_1^2-3 s_2 s_1+4 s_2^2\right) +$$
$$+s^2s_1^2 \left(s_1^2-49 s_2 s_1-2 s_2^2\right)
   +s\,s_1^3 \left(11 s_1^2+58 s_2 s_1+26 s_2^2\right) -s_1^4 \left(5 s_1^2+20 s_2 s_1+14 s_2^2\right)\bigr]+$$
$$+m^4\bigl[-20 s^5+63 s^4s_1 -4  s^3s_1
   \left(27 s_1-7 s_2\right)+
   4 s^2s_1 \left(37 s_1^2+32 s_2 s_1+5 s_2^2\right) -$$
   $$-2\,s\, s_1^2 \left(54 s_1^2+157 s_2 s_1+48 s_2^2\right) +s_1^3 \left(37 s_1^2+146 s_2 s_1+62
   s_2^2\right)\bigr]+$$
$$+2\,m^6 \bigl[27 s^4-46
   s^3s_1 -30 s^2s_1 \left(3 s_1+s_2\right) +2\,s\, s_1^2 \left(71 s_1+77 s_2\right)
   -s_1^2 \left(47 s_1^2+103 s_2 s_1+10 s_2^2\right)\bigr]+$$
$$+m^8\bigl[-56 s^3+63 s^2s_1 +s\left(6 s_1 s_2-238 s_1^2\right)
   +6 s_1 \left(18 s_1^2+9 s_2 s_1-s_2^2\right)\bigr]+$$
$$+2\,m^{10} \left[7 s^2-9 s s_1 +s_1 \left(9 s_2-13 s_1\right)\right]+m^{12}(12\,s+s_1)-6\,m^{14}\,.$$

The quantity $\overbrace{|\overline{M}_c|^2}$ can be derived from
the quantity$\overbrace{|M_c|^2}$ by the simple permutation
$s_1\leftrightarrows s_2.$ For the interference of the $M_c$ and
$\overline{M}_c$ amplitudes, which is symmetrical under the
permutation $s_1\leftrightarrows s_2$, we have
\begin{equation}\label{eq:B2}
\overbrace{M_c\overline{M}^*_c}=-\frac{\pi^2}{(s-m^2)^2\,s_1\,s_2}\Big\{2\,D_{c\bar c}\,L_a
-\frac{2\,B_{c\bar c}}{\lambda_1^5}\,L_1+
\end{equation}
$$\frac{s-m^2}{4s^2}\Big[c_1-\frac{12\,s\,m^2(s-m^2)}{\lambda_1^4}c_2-\frac{2}{\lambda_1^2}c_3 \Big] +(s_1\leftrightarrows s_2)\Big\}\,,$$
$$L_a=\ln\Big(\frac{\sqrt{a}+\sqrt{4m^2+a}}{2m}\Big)\,, \ \ a=s-s_{12}-m^2\,, $$

$$D_{c\bar c}=\frac{(s-s_{12})^3-m^2(3s^2-3ss_{12}+s_{12}^2)+m^4(4s_{12}-9s)+3m^6}{\sqrt{a(4m^2+a)}}\,,$$

$$B_{c\bar c}=(s-s_1)^5(s-s_{12})^2-m^2(s-s_1)\big[8s^5-6s^4(5s_1+2s_2)+s^3(38s_1^2+31s_1s_2+5s_2^2)-$$
$$-s^2s_1(15s_1^2+14s_1s_2+4s_2^2)-s\,s_1^2(4s_1^2+13s_1s_2+9s_2^2)+s_1^3(3s_1^2+8s_1s_2+5s_2^2)\big]+$$
$$+m^4\bigl[24 s^5-2 s^4\left(23 s_1+15 s_2\right) +2 s^3s_2 \left(22 s_1+5
   s_2\right) +
   s^2s_1 \left(41 s_1^2-15 s_2 s_1-5 s_2^2\right) +$$
$$+s\,s_1^2 \left(-19 s_1^2+18 s_2 s_1+16 s_2^2\right) -s_1^3 s_2 \left(13 s_1+8 s_2\right)\bigr]+m^6\bigl[-35 s^4+8 s^3\left(2 s_1+5 s_2\right) -$$
$$-2 s^2\left(13 s_1^2-3 s_2 s_1+5 s_2^2\right) +s\,s_1 \left(43 s_1^2-s_2 s_1+3 s_2^2\right)
   +s_1^2 \left(-13 s_1^2+7 s_2 s_1+s_2^2\right)\bigr]+$$
$$+m^8\left[25 s^3-15
   s^2\left(s_1+2 s_2\right) +s\left(-37 s_1^2-32 s_2 s_1+5 s_2^2\right) +s_1 \left(26 s_1^2+s_2 s_1-2 s_2^2\right)\right]-$$
$$-m^{10}\big[6s^2-6\,s(3s_1+2s_2)+18\,s_1^2-13\,s_1s_2+s_2^2\big]-2m^{12}(s+2s_1+s_2)+m^{14}\,,$$

$$c_1=s(6 s^2-10 s s_{12}+3 s_{12}^2)+m^2\Big[38\,s^2-11\,s\,s_{12}+5\,s_{12}^2+m^2(10\,s+3\,s_{12})-6\,m^4+\frac{4\,m^2(a^2-4\,m^2)}{s-m^2}\Big]\,,$$
$$c_2=s\big[3\,s^3-s^2(9s_1+4s_2)+s(9\,s_1^2+10\,s_1s_2+s_2^2)-3\,s_1\,s_{12}^2\big]+$$
$$+m^2\big[3\,s^3+s^2(-22\,s_1+4s_2)+2\,s(9\,s_1^2+8\,s_1s_2-s_2^2)-s_1\,s_{12}^2\big]+$$
$$+m^4\big(-11\,s^2-25\,s\,s_1+4\,s\,s_2+5\,s_1^2+6\,s_1\,s_2+s_2^2\big)+m^6\big[s-4(2s_1+s_2)\big]+4\,m^8\,,$$

$$c_3=s^2 (3 s-4 s_1) (s-s_{12}){}^2+$$
$$+m^2s\big[-7 s^3+s^2(30 s_2-28 s_1) +s\,(39 s_1^2+20 s_2 s_1-7 s_2^2) +s_1 (-5 s_1^2-4 s_1 s_2+s_2^2)\big]-$$
$$-m^4\big[69 s^3+s^2(70 s_1-2 s_2) - s\,(25 s_1^2+16 s_2 s_1+3 s_2^2)+s_1 s_{12}^2\big]+$$
$$+m^6 \big[39\, s^2-22\, s\,(2 s_1+s_2) +5\, s_1^2+6 s_1\, s_2+s_2^2\big]+m^8 \big[30\, s-4 (2 s_1+s_2)\big]+4 m^{10}\,.$$

The Borsellino-contribution is more complicated
$$\overbrace{|M_b|^2}=\frac{4\pi^2}{s-m^2}\Big\{\frac{L_1^{(b)}}{\lambda_1}+\frac{L_2}{2\,\lambda_2}-\frac{L_{12}}{\lambda_{12}}
-\frac{2\,B_{s_1}\,L_{s_1}}{(s-m^2)\sqrt{s_1^5(s_1-4m^2)}}-\frac{B_{s_2}\,L_{s_2}}{s_1^4\sqrt{s_2^5(s_2-4m^2)^5}}+$$
$$+\frac{D_b\,L_a}{s_1^4\sqrt{a^5(4m^2+a)^5}}-\frac{s-m^2}{2\,s_1^4s_2^2}\Big[b_1-\frac{96\,m^8(s-2s_1-m^2)^2}{(s_2-4m^2)^2}-\frac{8\,m^4\,b_2}{s_2-4m^2}
-\frac{6\,m^4\,b_3}{a^2}-$$
\begin{equation}\label{eq:B3}
-\frac{6\,m^4(s-m^2)^2(s-s_1+3m^2)}{(4m^2+a)^2}-\frac{2\,s_1^2s_2^2\,b_4}{(s-m^2)^2}+\frac{2\,s_1^2\,b_5}{s-m^2}+\frac{b_6}{2(4m^2+a)}+\frac{b_7}{2\,a}\Big]\Big\}\,,
\end{equation}
$$L_1^{(b)}=\ln\Big(\frac{s(s-s_1)-m^2(2\,s+s_1)+m^4-(s-m^2)\lambda_1}{2\,m\,s_1\sqrt{s}}\Big)\,, \ L_{s_{1,2}}=\ln\Big(\frac{\sqrt{s_{1,2}}+\sqrt{s_{1,2}-4\,m^2}}{2\,m}\Big)\,,$$

$$L_{12}=\ln\Big(\frac{s_{12}-2m^2-\lambda_{12}}{2\,m\,\sqrt{s}}\Big)\,, \ \
\lambda_{12}=\sqrt{s_{12}^2-4m^2(s+s_{12})+4\,m^4}\,,$$
where $L_2=L_1(s_1\to s_2)$ and
$$B_{s_1}=s_1\big[s^2+s_2^2+s_{12}^2-s(s_1+2s_2)\big]+m^2(4\,s\,s_{12}-s_1^2-10\,s_1s_2-4\,s_2^2)-m^4\big[8\,s-3(s_1+4s_2)\big]-8\,m^6\,,$$
$$B_{s_2}=-s_1^2 s_2^4 \bigl[2 s^2-2 s\left(s_1+2 s_2\right) +s_1^2+4 s_2^2+2 s_1 s_2\bigr]+$$
$$+2 m^2\,s_1 s_2^3 \bigl[4 s_1^3+5 s_2 s_1^2+26 s_2^2 s_1+8 s_2^3+s^2 \left(8 s_1-4 s_2\right)-s
   \left(7 s_1^2+14 s_2 s_1+4 s_2^2\right)\bigr]+$$
$$+2\,m^4s_2^2\bigl[4 s^3 s_2 +s^2\left(-18 s_1^2+24 s_2 s_1+8 s_2^2\right) +4 s s_1 \left(3
   s_1^2+9 s_2 s_1+10 s_2^2\right) -$$
   $$-s_1 \left(8 s_1^3+s_2 s_1^2+129 s_2^2 s_1+92 s_2^3\right)\bigr]-$$
$$-8\,m^6s_2\bigl[6 s^3 s_2 + s^2\left(-2 s_1^2+20 s_2 s_1+19 s_2^2\right)
  +s s_2 \left(3 s_1^2+24 s_2 s_1+4 s_2^2\right) +s_1 s_2 \left(3 s_1^2-65 s_2 s_1-95 s_2^2\right)\bigr]+$$
$$ + 4\,m^8s_2\left[40 s^3+4 s^2\left(12
   s_1+29 s_2\right) +s\left(-24 s_1^2+16 s_2 s_1+70 s_2^2\right) +s_2 \left(-81 s_1^2-348 s_2 s_1+4 s_2^2\right)\right]-$$
$$-8\,m^{10}\left[24 s^3+92 s^2 s_2 +2 s s_2 \left(49 s_2-8 s_1\right) +s_2 \left(-10 s_1^2-140 s_2 s_1+17 s_2^2\right)\right]+$$
$$+16\,m^{12} \left[36 s^2+62 s s_2 +s_2 \left(23 s_2-20 s_1\right)\right]-32\,m^{14}(18\,s+13\,s_2)+192\,m^{16}\,,$$

$$D_b=s_1^2(s-s_{12})^4(2s^2-4\,s\,s_{12}+3\,s_1^2+4\,s_2^2+6\,s_1 s_2)+$$
$$+2\,m^2s_1(s-s_{12})^3\bigl[10 s_1^3+34 s_2 s_1^2+34 s_2^2 s_1+8 s_2^3+2 s^2 \left(5 s_1+6 s_2\right)
   -s \left(19 s_1^2+44 s_2 s_1+20 s_2^2\right)\bigr]-$$
$$-2 m^4(s-s_{12})^2 \bigl[12 s^4-4 s^3\left(3 s_1+5 s_2\right) +s^2\left(-45 s_1^2-76 s_2 s_1+8 s_2^2\right) +$$
   $$+2 s s_1 \left(39 s_1^2+121 s_2 s_1+68 s_2^2\right) -s_1 \left(29 s_1^3+143 s_2 s_1^2+165 s_2^2 s_1+52 s_2^3\right)\bigr]-$$
$$-2\,m^6\bigl[16 s^5+4 s^4\left(s_1-5 s_2\right)-4 s^3\left(45 s_1^2+70 s_2 s_1+6 s_2^2\right) +s^2\left(370 s_1^3+960 s_2 s_1^2+604 s_2^2
   s_1+44 s_2^3\right) -$$
   $$-s\left(291 s_1^4+1021 s_2 s_1^3+1116 s_2^2 s_1^2+408 s_2^3 s_1+16 s_2^4\right) +s_1 \bigl(81 s_1^4+356 s_2 s_1^3+543 s_2^2 s_1^2+
  348 s_2^3 s_1+80 s_2^4\bigr)\bigr]+$$
$$+m^8\bigl[40 s^4-16 s^3\left(13 s_1+15 s_2\right) +s^2\left(438 s_1^2+592 s_2 s_1+256 s_2^2\right) -4 s\bigl(131 s_1^3+113 s_2 s_1^2+10 s_2^2 s_1
   +10 s_2^3\bigr) + $$
   $$+269 s_1^4-16 s_2^4-256 s_1 s_2^3-276 s_1^2 s_2^2+224 s_1^3 s_2\bigr]+$$
$$+2 m^{10}\left[96 s^3-24 s^2\left(s_1+3 s_2\right) -2 s\left(87 s_1^2+210 s_2 s_1+44 s_2^2\right) +75 s_1^3+28 s_2^3+388 s_1 s_2^2+462 s_1^2
   s_2\right]-$$
$$-2\,m^{12}\left[68 s^2-4 s\left(49 s_1+51
   s_2\right) +245 s_1^2+4 s_2^2+156 s_1 s_2\right]-8\,m^{14}\left(20 s+25 s_1+19 s_2\right)+120\,m^{16}\,,$$

$$b_1=-s_1^2\left(s^2-2 s_1 s+s_1^2+4 s_2^2\right)-2 m^2 s_1 \left(8 s^2-19 s s_1 +17 s_1^2+8 s_2^2\right)-$$
$$-m^4(16s^2-64s s_1+149 s_1^2)+16 m^6(2 s-7 s_1)-16 m^8\,, $$
$$b_2= s_1^2 \left(2 s_1-s\right) +m^2(8 s^2-44 s_1 s+61 s_1^2)+44 m^4 s_1-8m^6- \frac{16 m^2 s_1^2 \left(2 m^2+s_1\right) }{s-m^2}\,,$$
$$b_3=s^2+2 s_1^2-3 s s_1-m^2(2 s-3 s_1)+m^4\,, \ b_4=s_2^2+s_{12}^2-8 m^2 s_2+8 m^4\,,$$
$$b_5=s_2^2(s_1+2 s_2)+4 m^2(s_1^2+s_2^2)+24 m^4 s_1+32 m^6\,,$$
$$b_6=
   s s_1^2\left(s-s_1\right){}^2+m^2\left(2 s^4+8 s^3 s_1 -21 s^2 s_1^2 +16 s s_1^3 -5 s_1^4\right)+$$
   $$+m^4\left(16 s^3+96 s^2 s_1 -153 s s_1^2 +50 s_1^3\right)
-3 m^6\left(20 s^2-120 s_1 s+49 s_1^2\right)-48 m^8(4 s-s_1)+234 m^{10}\,,$$
$$b_7=
\left(s-s_1\right){}^2 s_1^2 \left(2 s_1-s\right)+m^2\left(2 s^4-24 s^3 s_1 +81 s^2 s_1^2 -116 s s_1^3 +73 s_1^4\right)-$$
$$-m^4(16 s^3-160 s^2 s_1+367s s_1^2-272 s_1^3)
+m^6(36 s^2-248 s s_1+287 s_1^2)-16 m^8(2 s-7 s_1)+10 m^{10}-$$
$$-\frac{16 m^2 s_1^4(s_1+2 m^2)}{s-m^2}\,.$$

To obtain $\overbrace{|\overline{M}_b|^2}$ we have to permutate $s_1$ and $s_2$ in $\overbrace{|M_b|^2}.$
The contribution of the interference $\overbrace{M_b}$ and $M_b$ is defined as
\begin{equation}\label{eq:B4}
\overbrace{M_b\,\overline{M}_b^*}=\frac{\pi^2}{s-m^2}\Big\{\frac{(s-s_1-5m^2)L_1+(s-s_1+3m^2)L_1^{(b)}}{\lambda_1(s-s_1-m^2)}
-\frac{2\,D_{b\bar{b}}\,L_a}{\sqrt{a^5(4m^2+a)}}-
\end{equation}
$$-\frac{1}{[a(s-m^2)+s_1s_2]}\Big[\frac{2\,\overline{B}_{s_1}L_{s_1}}{(s-m^2)s_2^2\sqrt{s_1^3(s_1-4m^2)}}-
\frac{\overline{B}_{12}\overline{L}_{12}}{\sqrt{a^5[a(s-m^2)^2+4m^2s_1\,s_2]}}\Big]+$$
$$+\frac{12m^4-8m^2(4m^2+a)+(4m^2+a)^2}{(s-m^2)s_1^2\,s_2^2\,a^2(4m^2+a)}C_{b\bar{b}}+(s_1\rightleftarrows s_2)\Big\}\,,$$
$$\overline{L}_{12}=\ln\Big[\frac{(s-m^2)\sqrt{a}}{2m\sqrt{s_1\,s_2}}\Big(1+\sqrt{1+\frac{4m^2s_1\,s_2}{a(s-m^2)^2}}\Big)\Big]\,,$$
$$C_{b\bar{b}}=2m^2(s-m^2)^2(s_1^2+s_2^2)-s_1\,s_2(4m^2+a)(s_1+s_2)^2\,,$$
$$D_{b\bar{b}}=-3 s^2+2 s s_{12}-2 (s_1^2-s_2 s_1+s_2^2)+23 m^4+4 m^2 s+\frac{1}{s_1 s_2(s-m^2)}\big[8 m^6(s_1-s_2)^2-$$
$$-4 m^4s_{12}(s_{12}^2-6 s_1 s_2)-2m^2\big((s_1-s_2)^4-12 s_1^2s_2^2\big)+s_{12}(s_1^4-6 s_1^3s_2-4s_1^2s_2^2-6 s_1 s_2^3+s_2^4)\big]-$$
$$-\frac{s^4-4 s^3m^2-6 m^4s^2+28m^6 s-19m^8}{a(s-m^2)+s_1 s_2}+$$
$$+\Big\{\frac{s^3-6 s_2^3+7 s s_2^2-4 s^2 s_2+m^2(-9 s^2+18 s_2 s-7 s_2^2)+3 m^4(5 s-2 s_2)+73 m^6}{2 s_1}+$$
$$+\frac{2 m^2\big[(s-s_2)^3-3 m^2(s-s_2)^2+m^4(s_2-9 s)+11 m^6\big]}{s_1^2}+\frac{1}{(s-m^2)(s-s_1-m^2)}\big[-38 m^8+$$
$$+m^6(56 s +62 s_2)-m^4(12 s^2+26 s s_2+5 s_2^2)-2m^2(4 s^3+3 s^2s_2+7 s s_2^2+6 s_2^3)+2(s+s_2)(s^3+2 s_2^3)+$$
$$+3 s^2s_2^2\big]+\frac{24 m^6\big[(s-s_1)(s-s_{12})+m^2(s_2-2s_1)-m^4\big]}{s_1^2(4m^2+a)}+(s_1\rightleftarrows s_2)\Big\}\,,$$

$$\overline{B}_{s_1}=-s_1 s_2 \bigl[s^3(s_1-s_2)+s^2(-2 s_1^2+s_2 s_1+2 s_2^2) +s(2 s_1^3-2 s_2 s_1^2-5 s_2^2 s_1-2 s_2^3)+2 s_1 s_2 s_{12}^2\bigr]+$$
$$+m^2\big[2 s^3 s_1(s_2-2 s_1)+s^2(4 s_1^3-s_2 s_1^2-5 s_2^2 s_1+4 s_2^3)-2 s s_2 (2 s_1^3-s_2 s_1^2+4 s_2^2 s_1+2 s_2^3)+$$
$$+s_1 s_2 (2 s_1^3+2 s_2 s_1^2+3 s_2^2 s_1+2 s_2^3)\big]+m^4\big[8 s^3 s_1+2 s^2(6 s_1^2-3 s_1 s_2-4 s_2^2)+$$
$$+s(-8 s_1^3+13 s_2 s_1^2+15 s_2^2 s_1-8 s_2^3) +s_2 (2 s_1^3+7 s_2 s_1^2+10 s_2^2 s_1+4 s_2^3)\big]+$$
$$+m^6\big[-40 s^2 s_1+s(-12 s_1^2+6 s_1s_2+32 s_2^2)+4s_1^3-11 s_1^2s_2-27 s_1 s_2^2 +4 s_2^3\big]+$$
$$+2 m^8(28 s s_1+2s_1^2-s_1 s_2-12s_2^2)-24 m^{10} s_1\,,$$

$$\overline{B}_{12}=s(s-s_{12})^2[s^2+(s-s_{12})^2]-m^2\big[10 s^4 -28 s^3 s_{12}+s^2(29 s_{12}^2+4 s_1 s_2)-$$
$$-2 s s_{12}(6 s_{12}^2+5 s_1 s_2)+s_{12}^2(s_{12}^2+6 s_1 s_2)\big]+$$
$$+m^4\big[28 s^3 -64 s^2 s_{12}+s(53 s_{12}^2+8 s_1 s_2)-2 s_{12}(8 s_{12}^2+5s_1 s_2)\big]-$$
$$-m^6(44s^2 -68 s s_{12}+31 s_{12}^2-12 s_1 s_2)+m^8(34 s-26 s_{12})-10 m^{10}\,.$$

The corresponding contribution of the interference between Borsellino and Compton-like diagrams reads
\begin{equation}\label{eq:B5}
\overbrace{M_c\,M_b^*}=\frac{2 \pi^2}{s_1(s-m^2)}\Big\{\frac{1}{(a+s_2)}\Big[\frac{B_{bc} L_1}{\lambda_1^3}+\frac{s_1 B^{(b)} L_1^{(b)}}{\lambda_1^3(s-m^2)}
+\frac{1}{s_1(s-m^2)}\Big(\frac{(s_2-2m^2)B_{s_2} L_{s_2}}{\sqrt{s_2(s_2-4 m^2)}}-
\end{equation}
$$-\frac{a D_{bc} L_a}{\sqrt{a(4 m^2 +a)}}\Big)\Big] +\frac{1}{2(s-m^2)}\Big[\frac{\widetilde{B}_{bc} L_2}{\lambda_2^3}+
\frac{\widetilde{B}_{12} L_{12}}{\lambda_{12}^3}\Big] -(4m^2+a-s_2)C_{bc} \Big\}\,,$$
$$C_{bc}=\frac{s_1}{2 \lambda_2^2}-\frac{s_1}{\lambda_1^2}+\frac{(s+s_1-m^2)[-s_1 s_{12}+2m^2(s+2 s_1)-2m^4]}{2 \lambda_2^2\,\lambda_{12}^2}\,,$$
$$B_{bc}=(4m^2+a-s_2)[(s-s_1)^3-m^2(3 s^2-2s s_1+s_1^2)+3m^4(s+3 s_1)-m^6]\,,$$
$$B^{(b)}=(4m^2+a-s_2)[(s-s_1)^3+m^2(s^2-4 s s_1+5 s_1^2)-m^4(5 s+ s_1)+3 m^6]\,,$$
$$B_{s_2}=s_1(3 s^2-4ss_1-6ss_2+3s_{12}^2+5s_2^2)+m^2(8s^2-6ss_1+4s_1^2-26s_1s_2)+m^4(35s_1-16s)+8m^6\,,$$
$$D_{bc}=s_1(5s^2-8ss_1-10ss_2+5s_{12}^2+3s_2^2)+m^2(8s^2-2ss_1-22s_1s_2)+m^4(29s_1-16s)+8m^6\,,$$
$$\widetilde{B}_{bc}=(s-s_1-2 s_2)(s-s_2){}^3+m^2\big[-8 s^3+s^2(3 s_1+13 s_2) +4 s s_1 s_2 -s_2^2 (3 s_1+5 s_2)\big]+$$
$$+m^4\big[14 s^2-3 s s_{12}+s_2(s_1+7s_2)\big]+m^6(-8 s+s_1-5 s_2)+m^8\,,$$
$$\widetilde{B}_{12}=s_{12}^3(-s+s_1+2 s_2)+m^2\big[2 s^2(3 s_1+s_2)-2 s s_{12}(2 s_{12}+s_2)-s_{12}^2(13 s_1+19 s_2)\big]+$$
$$+2 m^4\big[6 s^2+22 s s_1+26s s_2+5 s_{12}(6 s_{12}+s_2)\big]-2 m^6(44 s+57 s_1+59 s_2)+76 m^8\,.$$

To derive the quantity $\overbrace{\overline{M}_c\,\overline{M}_b^*}$ one has to make the permutation $(s_1\leftrightarrows s_2)$ in the right hand side of Eq.~(B5).

The contribution of the interference $\overline{M}_c$ and $M_b$ can be written as
\begin{equation}\label{eq:B6}
\overbrace{\overline{M}_c\,M_b^*}=\frac{\pi^2}{s_2(s-m^2)^2}\Big\{\frac{2 \widetilde{B}_{s_1} L_{s_1}}{\sqrt{s_1^3(s_1-4 m^2)}}
-\frac{2 \widetilde{B}_{s_2} L_{s_2}}{s_1^2\sqrt{s_2^3(s_2-4 m^2)^3}}-\frac{2 D_{b\bar{c}} L_a}{s_1^2\sqrt{a(4 m^2+a)}}-
\end{equation}
$$-\frac{B_{b\bar{c}} L_2}{\lambda_2^3}-\frac{\overline{B}^{(b)} L_1^{(b)}}{\lambda_1^3}+\frac{2 m^2 B_{12} L_{12}}{\lambda_{12}^3}+C_{b\bar{c}}\Big\}\,,$$

$$C_{b\bar{c}}=\frac{s-m^2}{2 s}\Big\{-s+s_1-m^2+\frac{(s-s_1-m^2)(s^2-s s_{12}-s_1 s_2+m^2 s_{12}-m^4)}{\lambda_1^2}+$$
$$+\big[s(s_2-s_1)-s_2 s_{12}+m^2(2 s +s_1+3 s_2)-2 m^4\big]\Big(\frac{2 m^2}{\lambda_{12}^2}+\frac{s+s_2-m^2}{\lambda_2^2}\Big)\Big\}+$$
$$+\frac{2}{s_2}\Big\{(s-s_{12})^2-2 m^2(s-s_{12})-3 m^4+\frac{1}{(s_{12}-4 m^2)^2}\Big[-\frac{24 m^6(s-m^2)(s-2 s_{12}+7m^2)}{(s_2-4 m^2)}+$$
$$+\frac{1}{s_1}\big[(s-s_{12})s_{12}^4+m^2s_{12}^2(2 s^2 -4 s s_{12}+7 s_{12}^2)-8 m^4s_{12}(2 s^2+s s_{12}+s_{12}^2)+m^6[8 s^2+48 s s_{12}-26 s_{12}^2]-$$
$$-16m^8(s-2s_{12})+8m^{10}\big]\Big]-$$
$$-\frac{2 m^2(s-m^2)^2(s_{12}^2-4 m^4)}{s_1^2(s_{12}-4 m^2)}-\frac{s-m^2}{\lambda_2^2}\big[s(s-s_2)(s-s_{12})+m^2(s^2+2ss_1-6ss_2+3s_2 s_{12})-$$
$$-m^4(5 s +s_1+8 s_2)+3 m^6\big]\Big\}\,,$$
$$\widetilde{B}_{s_1}=s_1(s^2+2 s s_{12}+s_{12}^2)+2 m^2s_{12}(2 s-s_1-2s_2)+m^4(8 s-13 s_1-4 s_2)+8 m^6\,,$$
$$\widetilde{B}_{s_2}=s_1s_2^3[(s-s_{12})^2+s_2^2]-2m^2s_2^2[3s^2s_1-s(5s_1^2+3 s_1 s_2+2 s_2^2)+s_1(2 s_1^2+6 s_1 s_2+5 s_2^2)]+$$
$$+m^4s_2[8 s^2s_1-4 s(2 s_1^2-3 s_1s_2+4 s_2^2)+s_2(6 s_1^2+s_1 s_2-4 s_2^2)]+2m^6s_2[8 s^2-8 s(4 s_1+s_2)+20 s_1^2+$$
$$+21 s_1s_2+8 s_2^2]+8 m^8[4 s^2+4s s_2+s_2(2 s_2-s_1)]-16 m^{10}(4 s +3 s_2)+32 m^{12}\,,$$
$$B_{b\bar{c}}=(s-s_2)[(s^2-s_2^2)(s_1+2 s_2)-2 s s_1 s_2]-m^2[4 s^3+s^2(3 s_1-2 s_2)+2 s s_2(s_2-5 s_1)+s_2^2(9 s_1+8 s_2)]+$$
$$+m^4[12 s^2+3 s(s_1-6 s_2)+s_2(9 s_1+20 s_2)]-m^6(12 s+s_1+18 s_2)+4 m^8\,,$$
$$\overline{B}^{(b)}=(s-s_1)^3(s_1+2s_2)+m^2[4 s^3-s^2(11 s_1+2 s_2)+4 s s_1(s_1-s_2)+s_1^2(3 s_1+10 s_2)]-$$
$$-m^4(4 s^2+13 s s_1+2 s s_2+s_1^2+6 s_1 s_2)+m^6(-4 s-9 s_1+2 s_2)+4 m^8\,,$$
$$B_{12}=s_{12}(2s s_1+3 s s_2-2 s_1^2-3 s_1 s_2-s_2^2)-m^2(10 s^2+8 s s_2-14 s_1^2-23 s_1 s_2-9 s_2^2)-$$
$$-4 m^4(3 s+8 s_1+6 s_2)+22 m^6\,,$$
$$D_{b\bar{c}}=s_1(s^3-3 s^2 s_{12}+4 s s_{12}^2-2 s_{12}^3)+m^2[s^2(4 s_2-5 s_1)+2 s(4 s_1^2+ s_1 s_2-2 s_2^2)-2 s_1^2(2 s_1+s_2)]+$$
$$+m^4(8 s^2-17 s s_1-8 s s_2+19 s_1^2+9 s_1 s_2+ 4 s_2^2)+m^6(5 s_1+4 s_2)-8 m^8\,.$$

In accordance with Eq.~(10) and definition of quantity $\overbrace{W}$ given in this Appendix, we can write
$$\frac{d \sigma}{d\,s_1\,d\,s_2}=\frac{\alpha^3}{2\,\pi^2(s-m^2)}\overbrace{\sum_{pol}|M|^2}\,,$$
where $\sum_{pol}|M|^2$ is defined by Eq.~(9).



%

\end{document}